\documentclass[superscriptaddress,showpacs,preprintnumbers,amsmath,amssymb,floatfix,elsart,prd]{revtex4}
\usepackage{graphicx}
\usepackage{dcolumn}
\usepackage{bm}
\usepackage{color}

\newcommand{\beq}{\begin{equation}}
\newcommand{\eeq}{\end{equation}}
\newcommand{\bey}{\begin{eqnarray}}
\newcommand{\eey}{\end{eqnarray}}

\begin{document}

\title{An analytical anisotropic compact stellar model of embedding class I}

\author{Lipi Baskey}
\email{lipibaskey@gmail.com} \affiliation{Department of Mathematics, Govt. General Degree College at Kushmandi,\\
Dakshin Dinajpur 733121, West Bengal, India}

\author{Shyam Das}
\email{dasshyam321@gmail.com} \affiliation{Department of Physics, P. D. Women's College,\\
Jalpaiguri 735101, West Bengal, India}

\author{Farook Rahaman}
\email{rahaman@iucaa.ernet.in} \affiliation{Department of
Mathematics, Jadavpur University, Kolkata 700032, West Bengal,
India}

\date{\today}

\begin{abstract}
A class of solutions of Einstein field equations satisfying Karmarkar embedding condition is presented which could describe static, spherical fluid configurations, and could serve as models for compact stars. The fluid under consideration has unequal principal stresses i.e. fluid is locally anisotropic. A certain physically motivated geometry of metric potential has been chosen and codependency of the metric potentials outlines the formation of the model. 
 The exterior spacetime is assumed as described by the exterior Schwarzschild solution. The smooth matching of the interior to the exterior Schwarzschild spacetime metric across the boundary and the condition that radial pressure is zero across the boundary lead us to determine the model parameters. Physical requirements and stability analysis of the model demanded for a physically realistic star are satisfied. The developed model has been investigated graphically by exploring data from some of the known compact objects. The mass-radius ($M-R$) relationship that shows the maximum mass admissible for observed pulsars for a given surface density has also been investigated. Moreover, the physical profile of the moment of inertia ($I$) thus obtained from the solutions is confirmed by the
Bejger-Haensel concept.
\end{abstract}

\maketitle

\section{Introduction}
\label{sec:intro}

Theoretical modeling of relativistic stellar structure has become a trend since the attainment of solutions of Einstein field equations (EFEs) by Karl Schwarzschild for the spherically symmetric static exterior \cite{ks1} and interior \cite{ks2} of a stellar object. Schwarzschild obtained the solutions considering a matter distribution with uniform density \cite{ks1,ks2}. Later, the works of Tolman \cite{tolman} and Oppenheimer and Volkoff \cite{OV} branched out the study for more realistic modeling in the proximity of data-based facts as Tolman mentioned eight different exact solutions for EFEs while in the same issue of Physical Review, Oppenheimer and Volkoff were first to obtain computational solutions of EFEs for a degenerate neutron gas. Initially, it was assumed that the nature of spherically symmetric matter is similar to that of a perfect fluid, where radial pressure coincides with tangential pressure. But in 1922, Jeans \cite{jeans} changed this concept suggesting that due to the extreme and unusual conditions reigning through the interior of compact objects, anisotropy needs to be given the importance of to study its nature of matter distribution. In physics, the term anisotropy is used to describe the direction-dependent properties of materials. However, in context of compact stars anisotropy denotes the difference of radial and tangential pressures. Eleven years later Leimatre \cite{leimatre} modeled first anisotropic model entirely with tangential pressure and constant density. In 1972, Ruderman \cite{ruderman} proposed that highly compact objects are, in fact, anisotropic in nature predominantly because of its high density ($> 10^{15}gm/cc$). Bowers and Liang \cite{BL} worked on the anisotropic sphere in General Relativity. Dev and Gleiser \cite{DG1,DG2} showed the significance of physical parameters like mass, structure etc. on anisotropy. Anisotropy can exists for various reasons such as presence of superfluid, solid core \cite{KW}, mixture of different fluids and viscosity \cite{HS}, phase transitions \cite{sokolov}, meson condensation \cite{sawyer}, strong magnetic field \cite{weber}, superconductivity \cite{migdal} to name a few. Modeling of anisotropic compact objects under various conditions with some certain assumptions are performed by many researchers like assuming barotropic equation of state \cite{MK07,RBBU}, assuming pressure anisotropy in (3 + 1)D spacetime \cite{bhar1,bhar2,bhar3}, assuming quadratic equation of state for stellar interior \cite{BM}, assuming quadratic envelope \cite{TMM} and so on. Herrera along with other collaborators \cite{herrera92,HPI,HOP,HSW} studied and analyzed the stability of the self-gravitating system in the presence of local anisotropy. A new class of exact solutions of EFEs for the anisotropic stellar model has been pursued by Mak and Harko \cite{MH03}. Additionally, the duo has done some fascinating work with some other researchers in the context of anisotropic fluid matter distribution \cite{HM1,HM2,HM3,HM4,MDH}. Effect of anisotropy on the formation of the structure of a spherically symmetric stellar model has thoroughly been discussed on various literature \cite{MGRD,MBH,MBG,DCRRG,KRMH,MMKP,TRSD}. Das et. al \cite{DRB} have investigated the diverse physical aspects to study the comportment of the compact stellar model
in anisotropic fluid matter distribution by performing a comparative study of the proposed model with observational data.\\
 The general approach for modeling a relativistic compact stellar object is to particularize gravitational potentials, matter variables, consider a specific type of equation of state or to make use of geometric constraints on specific spacetime geometry. Some examples can be provided as Feroze and Siddiqui \cite{FS} and Malaver have considered the linear equation of state \cite{malaver1} and quadratic equations of state \cite{malaver2,malaver3} for the matter distribution and specified particular forms for the gravitational potential and electric field intensity. Considering a polytropic equation of state Mafa Takisa and Maharaj \cite{TM} have obtained new exact solutions to the Einstein-Maxwell system of equations and Thirukkanesh and Ragel \cite{TR} have obtained anisotropic fluid model by solving Einstein field equations. Another way of describing a compact stellar model in relativistic field equation is to embed a $n$ dimensional Euclidean space. Embedding of $n$ dimensional space $V_{n}$ into a $n+p$ dimensional Euclidean space $E_{n+p}$ catches the eye of the researchers after the work on brane theory by Randall and Sundrum \cite{RS}. Though, from the mathematical aspect, the theory of embedding starts with the definition of manifolds by Riemann. Eventually, L. Schlaefli \cite{schlai} laid the foundation in 1871 when he conjectured that spacetime can be embedded into higher dimensional pseudo-Euclidean space. Later, it was proven by Janet and Cartan \cite{janet,cartan}, famously known as Janet-Cartan theorem, that to embed any $4$-dimensional space-time, both locally and isometrically, maximum $10$-dimensions are needed. Subsequently, it was extended to manifold with an indefinite metric by Friedman \cite{friedman}.  Recently an intriguing scheme to model stable configuration is to embed a 4-dimensional spacetime into higher dimensional space though another particular form of Buchdahl metric is also studied by Vaidya and Tikekar \cite{VT} and Tikekar \cite{tikekar} by embedding a 3-hyperspace into 4-dimensional space. Recently Buchdahl metric has been analyzed in spherically symmetric spacetime in terms of embedding by Singh et. al \cite{SPG} and in terms of Vaidya-Tikekar and Finch-Skea model by Maurya et al \cite{MBJKPP}. Condition for embedding a 4-dimensional spacetime metric into 5-dimensional Euclidean space was first derived by K. R. Karmarkar \cite{karmarkar}, which is named after him as Karmarkar condition. In this condition, two metric potentials need to be dependent on each other. The simple expression between the metric potentials allows the researcher to choose general forms of metric potential having physical viability to generate acceptable compact models. Any solution of EFEs satisfying Karmarkar condition is considered to be of Class I. Numerous researchers have devoted their time in the modeling of both charged and uncharged stars using Karmarkar condition. Some of the recent literature backing the modeling of anisotropic compact stars using embedding class I condition are authored by Pandya et al \cite{PTGRS,PT}, Singh et al \cite{SBMP,SBLR,SERD,SMERD}, Gedela et al \cite{GBP1,GBP2,GPBP} and many more \cite{OMES,JMA,SSRSS,SSSR,PKMD,FMS,SF,TF,RSERD,MOJ,OML}. Recently, Govender et al \cite{GMSP} have studied the gravitational collapse of a spherically symmetric star by employing Karmarkar condition. It is worth mentioning that Schwarzschild exterior solution is of class II and the interior solution is of class I. The solution for the isotropic fluid sphere that satisfies Karmarkar condition is either Schwarzschild interior solution \cite{ks1,ks2} in which inner solution is conformally flat depicting limited configuration or Kohler-Chao \cite{KCN} solution for which inner solution is conformally non-flat depicting limitless configuration, yet is considered to obtain a new class of relativistic solutions.\\
{Maurya et al \cite{MGTR} have investigated new solutions for EFEs using Karmarkar condition considering a specific form of metric potential viz. $e^\lambda = 1 + \frac{(a - b)r^2}{1 + b r^2}$, with $a \neq b$. In this paper, by utilizing the special case of this metric we have studied some new features of compact stars which was not discussed in their work.} We have studied a solution of EFEs assuming a specific form of metric potential $e^\lambda = \frac{2(1+A r^2)}{2 -A r^2}$, $A$ being the constant parameter and $r$ being the radial coordinate and hence delved into a relativistic model of an anisotropic compact star of embedding class I in static, symmetric and spherical geometry. {Our metric can easily be obtained by substituting $a = A$ and $b = -{A \over 2}$ in the metric considered by \cite{MGTR}.} Since Buchdahl metric \cite{buchdahl59} is one such metric which satisfies all the criteria for physical acceptability of a stellar structure as listed by Delgaty and Lake \cite{DL}, so we have considered ansatz which is a special form of Buchdahl metric. The model is shown to execute linear equation of state. The estimated mass and radius are revealed to be as similar as the observed data for the stars $4$U$1820-30$, SMC X$-1$, SAX J $1808.4-3658$, $4$U$1608-52$, PSR J $1903 + 327$, Vela X$-1$ and $4$U$1538-52$. {Additionally, we have also studied the mass-radius relationship and radius-central density relationship for the chosen metric. Comparing the obtained result for a slow rotating configuration, we have also studied the mass-moment of inertia relationship for our prescribed model.}\\
Our paper has been organized as follows: In Sec.~\ref{sec2}, we have presented the basic equations governing the anisotropic system. In Sec.~\ref{sec3}, we have recalled Karmarkar condition and our proposed model have been discussed mathematically in Sec.~\ref{sec4}. Utilizing the matching condition on the boundary of the stellar object in Sec.~\ref{sec5} we have established a general form of the constant parameters and hence examined physical properties viz. regularity condition, energy condition, generalized Tolman-Oppenheimer-Volkoff (TOV) equation, causality condition, stability condition etc of several compact stars in Sec.~\ref{sec6} in the framework of general relativity. Finally, we have provided a short discussion of matter variables considering several stars in Sec.~\ref{sec7} and concluded our results in Sec.~\ref{sec8}.
\section{\label{sec2}Einstein field equations}
We consider the line element in $4$D co-ordinate system $(t,~r,~\theta,~\phi)$ to describe the interior of a static and spherically symmetric stellar configuration as,
\begin{equation}
ds^{2}=e^{\nu(r)}dt^{2}-e^{\lambda(r)}dr^{2}-r^{2}\left(d\theta^{2}+\sin^{2}\theta d\phi^{2} \right), \label{le}
\end{equation}
where the metric potentials $e^{\nu(r)}$ and $ e^{\lambda(r)} $ or more precisely $\nu(r)$ and $\lambda(r)$ are functions of the radial coordinate `$r$' only.\\
Assuming $8\pi G =~c~=~1$ the Einstein field equations can be written as,
 \begin{equation}
 G_{\mu \nu} = - T_{\mu \nu} = \left(R_{\mu \nu}-{1\over 2}R~g_{\mu \nu}\right), \label{tensor}
\end{equation}
where $G_{\mu \nu},~ T_{\mu \nu},~ R_{\mu \nu}, ~ g_{\mu \nu}$ and $R$ are Einstein tensor, the stress energy tensor, Ricci tensor, metric tensor and Ricci scalar respectively.\\
For an anisotropic matter distribution, the energy momentum tensor can be written as,
\begin{equation}
T_{\mu \nu} =  \left(\rho(r)+p_t(r)\right)U_\mu U_\nu-p_t(r) g_{\mu \nu}+\left(p_r(r)-p_t(r)\right)\chi_\mu \chi_\nu, \label{tensor2}
\end{equation}
where $\rho(r)$ is the energy density, $p_r(r)$ and $p_t(r)$ represent pressures along the radial and the transverse directions of the fluid configuration respectively. $U^\mu$ is the $4$-velocity and $\chi^\mu$ is an unit $4$-vector along the radial direction. The quantities obey the relations, $\chi_\mu \chi^\mu = 1$ and $\chi_\mu U^\mu = 0$.\\
The Einstein field equations (\ref{tensor}) governing the evolution of the system read as the following form for the metric (\ref{le}) along with the energy tensor (\ref{tensor2}),
\begin{eqnarray}
\rho(r)&=& \frac{1-e^{-\lambda}}{r^{2}}+\frac{e^{-\lambda}\lambda'}{r},  \label{dens}\\
 p_{r}(r)&=& \frac{e^{-\lambda}-1}{r^{2}}+\frac{e^{-\lambda}\nu'}{r},  \label{prs}\\
 p_t(r)&=& {e^{-\lambda} \over 2} \left(\nu''+ \frac{\nu'-\lambda'}{r} + \frac{\nu'^{2}-\nu'\lambda'}{2} \right), \label{prt}
\end{eqnarray}
where `prime' in Eqs.~(\ref{dens})-(\ref{prt}) denotes differentiation with respect to radial co-ordinate $r$. Also, the anisotropic factor is defined  as  $\Delta(r) = \left(p_t(r) - p_r(r)\right)$ and its expression for the stellar system is
\begin{equation}
\Delta(r)= \frac{e^{-\lambda}}{2}\left( \nu'' +\frac{\nu'^{2}}{2}-\frac{\nu'\lambda'}{2}- \frac{\nu'-\lambda'}{r} \right) + \frac{1- e^{-\lambda}}{r^2}.
\label{eqdelta}
\end{equation}

\begin{figure}[!htbp]
\centering 
\includegraphics[width=.49\textwidth]{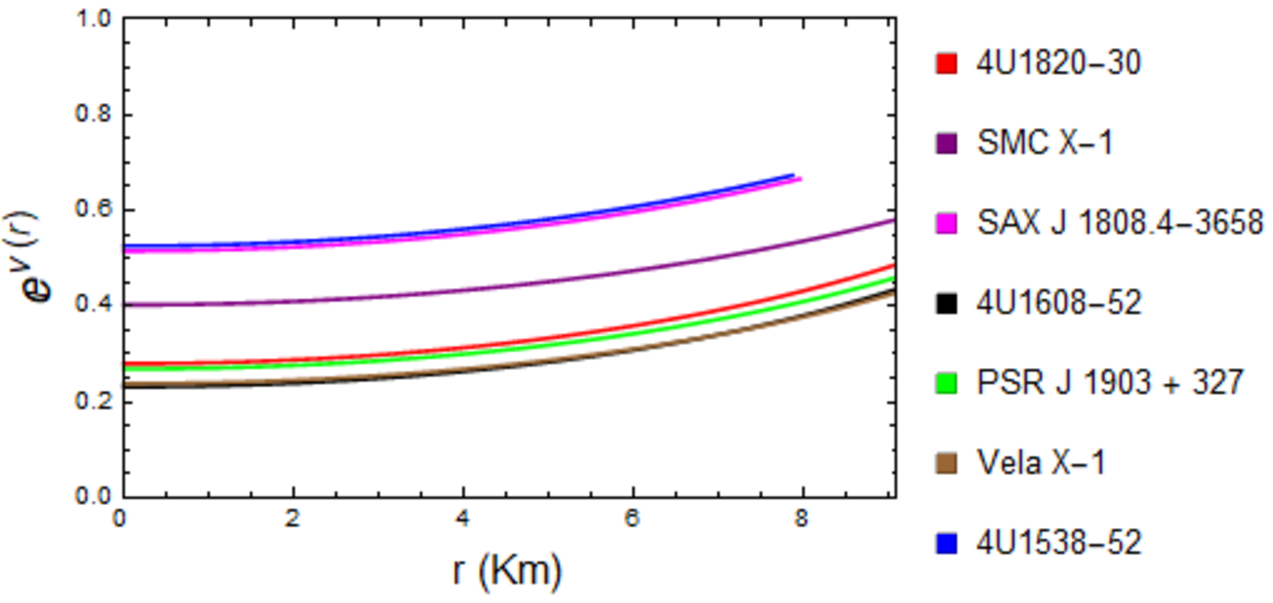}
\hfill
\includegraphics[width=.49\textwidth]{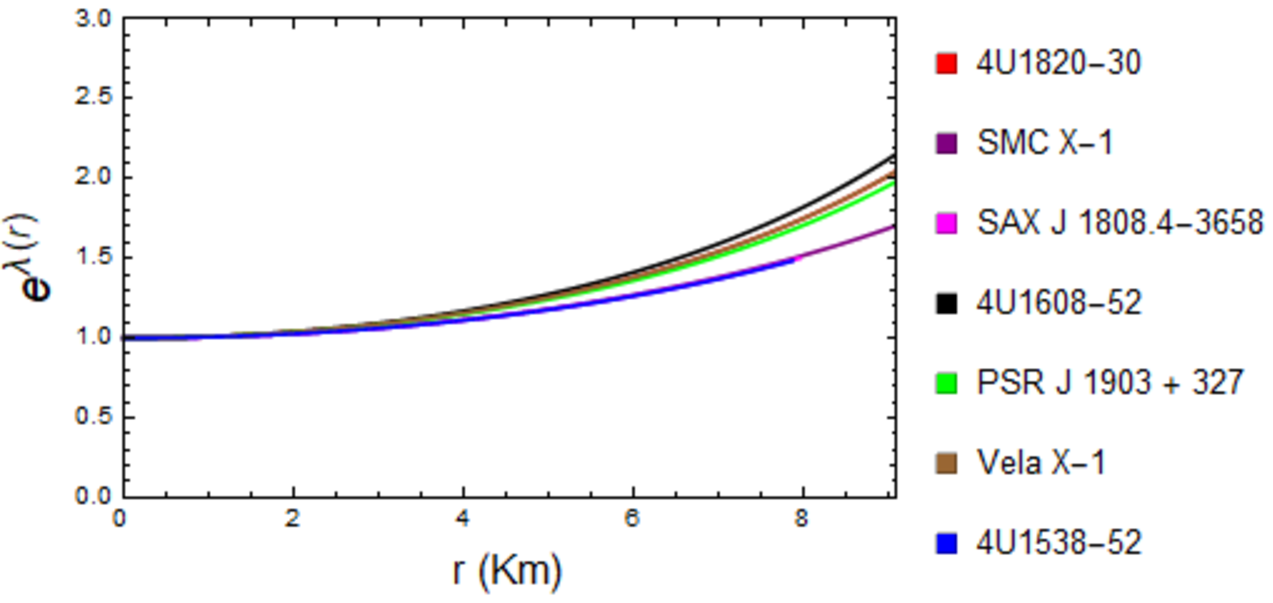}
\caption{\label{figmp} Behavior of metric potentials $e^{\nu}$ (left) and $e^{\lambda}$ (right) with respect to the radial coordinate $r$ for the compact stars corresponding to the numerical value of constants given in Table~\ref{table1}.}
\end{figure}
\section{\label{sec3}Karmarkar Condition}
The general theory of  relativity tells us that whenever an $n$ dimensional spacetime is embedded in a Pseudo Euclidean spacetime of $n+p$ dimension then `$p$' is called embedding class. A symmetric tensor $h_{\alpha\beta}$ of $4$ dimensional Riemannian space can be embed into a $5$ dimensional Pseudo Euclidean space if it is satisfies Gauss \cite{gauss} and Codazzi \cite{codazzi} condition and it can be written as,
\begin{eqnarray}
R_{\alpha \beta i j} &=& \epsilon \left(h_{\alpha i}h_{\beta j}-h_{\alpha j}h_{\beta i}\right),\label{eq8}\\
h_{\alpha \beta ; i}- h_{\alpha i ; \beta} &=& 0,\label{eq9}
\end{eqnarray}
where $R_{\alpha \beta i j}$ denotes curvature tensor.

Here, $\epsilon = 1$, when normal to the manifold is spacelike,\\
  $\epsilon = -1$, when normal to the manifold is timelike and the symbol `;' represent covariant derivative. Earlier Kasner \cite{kasner} investigated in the year $1921$ that a $4$ dimensional spacetime of spherically symmetric object can always be embedded in $6$ dimensional Pseudo Euclidean space and later Gupta and Goyel ($1975$) \cite{GG} have shown the same result with another coordinate transformation. In $1924$, Eddington \cite{eddington} found that an n-dimensional spacetime can always be embedded in m-dimensional Pseudo Euclidean space with $m=n(n+1)/2$ and to embed, the minimum extra dimension required is less than or equal to the number $(m-n)$ or same as $n(n-1)/2$. Therefore $4$ dimensional spherically symmetric line element Eq.~(\ref{le}) is of embedding class II.
From Eq.~(\ref{le}), the components of Riemann curvature tensor are expressed as,
\begin{eqnarray}
R_{1414}=-e^\nu \left(\frac{\nu''}{2}+\frac{\nu'^2}{4}-\frac{\lambda'\nu'}{4}\right),~~R_{1212}= \frac{r \lambda'}{2},\nonumber \\
R_{2323}= e^{-\lambda} r^2 {\sin^2\theta}(e^\lambda -1),~~R_{2424}= \frac{1}{2} \nu' r e^{\nu-\lambda},\nonumber \\
R_{3434}= R_{2424} \sin^2\theta ,~~R_{1224}=0, \nonumber \\
R_{1334} = R_{1224}\sin^2 \theta =0. \label{eq10}
\end{eqnarray}

In $1948$, K. R. Karmarkar derived a condition known as Karmarkar condition which allows us to embed any $4$ dimensional spacetime into $5$ dimension flat space.
Now the non zero components of the tensor $h_{\alpha i}$ corresponding to Eq.~(\ref{le}) are $h_{11}$, $h_{22}$, $h_{33}$, $h_{44}$ and $h_{14}(=h_{41})$ due to its symmetric nature and $h_{33}=h_{22}\sin^2\theta$.\\
Using these aforementioned components Eq.~(\ref{eq8}) reduced to
\begin{equation}
R_{1414}R_{2323} = R_{1212}R_{3434} +R_{1224}R_{1334},
\label{eq11}
\end{equation}
which is the expression for Karmarkar Condition. Later in $1981$, Sharma and Pandey \cite{PS} clarified that condition (\ref{eq11}) is only necessary condition for a class one spacetime to be $4$ dimensional spacetime but it is not sufficient. In order to be a class one, a spacetime must satisfies Eq.~(\ref{eq11}) along with $R_{2323}\neq 0$.

On substituting all the values of Eq.~(\ref{eq10}) in Eq.~(\ref{eq11}), we obtain the following differential equation,
\begin{equation}
\frac{2\nu''}{\nu'}+ \nu' = \frac{\lambda' e^\lambda}{e^\lambda - 1},
\label{eq12}
\end{equation}
with $e^\lambda \neq 1$.
Solving Eq.~(\ref{eq12}) we obtain the relationship between $\lambda$ and $\nu$ as,
\begin{equation}
e^{\nu(r)} = \left[C + D \int{\sqrt{e^{\lambda(r)}-1}}dr \right]^2,
\label{nu}
\end{equation}
where $C$ and $D$ being non zero integrating constants. The distinctive feature of class I spacetime is condition (\ref{nu}) i.e. the co-dependency of the metric potentials which further provides scope to generate anisotropic model of embedding class I by specifically choosing one of the metric potentials. \\
According to Maurya et. al \cite{MGRC}, anisotropic factor becomes,
\begin{equation}
\Delta(r) = {\nu'(r) \over 4e^\lambda} \left( \frac{\nu'(r)e^\lambda}{2 r D^2} -1\right) \left( {2 \over r} - \frac{\lambda'(r)}{e^\lambda - 1}\right).
\label{eqanim}
\end{equation}
Clearly, pressure anisotropy will vanish if either one of the term on RHS (right hand side) of Eq.~(\ref{eqanim}) will become zero. Now if the term $\left( \frac{\nu'(r)e^\lambda}{2 r D^2} -1\right)$ vanishes then it leads to Kohler-Chao solution and if $\left( {2 \over r} - \frac{\lambda'(r)}{e^\lambda - 1}\right)$ vanishes then it yields Schwarzschild interior solution.
\section{\label{sec4}A particular model}
The fundamental approach of theoretical modeling is to find the exact solutions of the system of equations represented by Eqs.~(\ref{dens})-(\ref{prt}) thus leading to determine the structure of spacetime for anisotropic fluid distribution. Clearly, if $p_r = p_t$ then the above system of equations lead to a perfect fluid like matter distribution. Now the EFEs are consist of three equations with five unknowns namely
$\lambda(r),~\nu(r),~\rho(r),~p_r(r)$ and $p_t(r)$. However, Karmarkar condition furnish a relation between the two metric potentials $\lambda(r)$ and $\nu(r)$, providing total four equations with five unknowns. To balance this system of equation we consider the metric potential which is a special form of Buchdahl ansatz \cite{buchdahl59}. Specifically this ansatz has been considered by Maurya et. al \cite{MGDJA} and it is given as
\begin{equation}
e^{\lambda(r)} = \frac{2(1 + A r^2)}{2 - A r^2},
\label{etlamda}
\end{equation}
where $A$ is a non negative constant.
It is also to be noted that as $A = 0$ makes the metric function flat as $e^\lambda=1$, so clearly $A \neq 0$ making $A$ strictly positive.
Besides the regularity, known singularity and the fact that metric function is finite at the center of the star ($r = 0$) satisfies the basic physical requirement of a compact star thus making it a physically tenable model.

Plugging Eq.~(\ref{etlamda}) onto Eq.~(\ref{nu}) we have,
\begin{equation}
e^{\nu(r)}=\left[ C - \frac{D\sqrt{3(2 - A r^2)}}{{\sqrt{A}}}\right]^2.
\label{etnu}
\end{equation}
Also the function $e^\nu$ needs to be finite and positive at the center. Observing Fig.~\ref{figmp} it can be concluded that $e^\nu$ is monotonically increasing with radial coordinate `$r$' throughout the star, implying $e^\nu$ may produce a metric potential for a  viable model as proposed by Lake \cite{lake}.

The expressions for energy density, radial pressure and transverse pressure are thus obtained as following,
\begin{eqnarray}
 \rho(r)&=& \frac{3 A(3 + A r^2)}{2 (1 + A r^2)^2},  \label{density}\\
 p_{r}(r)&=& \frac{A \left(3 C \sqrt{\frac{A }{2 - A r^2}} - 5\sqrt{3} D\right)}{2 (1 + A r^2)\left(\sqrt{3}D  - C \sqrt{\frac{A }{2 - A r^2}}\right)},  \label{rp}\\
 p_t(r)&=& \frac{A \left(3 C \sqrt{\frac{A }{2 - A r^2}} - \sqrt{3} D  (5 + A r^2)\right)}{2 (1 + A r^2)^2\left(\sqrt{3} D  - C \sqrt{\frac{A }{2 - A r^2}} \right)}. \label{tp}
\end{eqnarray}

The anisotropy becomes,
\begin{equation}
\Delta(r)= \frac{3 A r^3}{4 (1 + A r^2)}.
\label{aniso}
\end{equation}
Also the mass function and the compactness factor are given as,
\begin{eqnarray}
m(r)&=& 4 \pi \int_{0}^{r}  \rho(\omega) \omega^2 d\omega = \frac{3 A r^3}{4(1 + A r^2)}, \label{massf}\\
u(r) &=& \frac{m(r)}{r}= \frac{3 A r^2}{4(1 + A r^2)}. \label{compact}
\end{eqnarray}
\section{\label{sec5}The Boundary Conditions}
To analyze a compact object in general theory of relativity it is important to model the spacetime insofar two distinct manifolds are unified at a common boundary. These junction conditions determine the constants for the anisotropic fluid configuration i.e. $C$, $D$ and $A$ in this case and these are given as,\\
\textbf{(i)} Continuity of the first fundamental form at the boundary i.e. the interior solution should be matched to the vacuum exterior Schwarzschild solution at the boundary $r = R$ of the star known as the radius of the star. Generally the process executed here is Darmois-Israel formalism  based on Gauss-Codazzi decomposition of spacetime. It expresses the surface properties in terms of jump of extrinsic curvature across the boundary as the function of boundary's intrinsic coordinates \cite{MK}. Israel \cite{israel} formulated his work considering the idea that the $4$ dimensional coordinates may be chosen independently on both side of the boundary. In his innovating work Darmois \cite{darmois} first calculated that the boundary of the structure is to be considered as the periphery of two different manifolds glued together at this boundary.\\
\textbf{(ii)} Continuity of the second fundamental form at the boundary i.e. the radial pressure must vanish at the boundary. Mathematically $p_r(r = R)=0$ \cite{MS}.\\
The exterior spacetime on Schwarzschild metric is expressed as,
\begin{equation}
ds^2 = \left( 1- \frac{2M}{r} \right) dt^2 - \left( 1 -\frac{2M}{r}\right)^{-1}dr^2 - r^2 (d\theta^2 +\sin^{2}\theta d\phi^{2}),
\label{bcs}
\end{equation}
at $r > 2M$, $M$ being stellar mass.
The  advancement of metric function over the limiting surface i.e. at the boundary yields,
\begin{eqnarray}
e^{\nu(R)} &=& \left(1 - \frac{2 M}{r}\right)|_R,  \label{nu1}\\
e^{-\lambda(R)} &=& \left( 1 - \frac{2 M}{r}\right)|_R. \label{lamda1}
\end{eqnarray}

Utilizing above equations with Eqs.~(\ref{etnu}) and (\ref{etlamda}) respectively we get,
\begin{eqnarray}
\left( C - \frac{D \sqrt{3(2 - A R^2)}}{\sqrt{A}}\right )^2 &=& 1 - \frac{2 M}{R},  \label{nu2}\\
\frac{2 - A R^2}{2(1 + A R^2)} &=& 1 - \frac{2 M}{R}. \label{lamda2}
\end{eqnarray}

Again the later condition $p_r(r = R)= 0$ gives,
\begin{equation}
A \left(3 C \sqrt{\frac{A}{2 - A R^2}} - 5\sqrt{3} D \right) = 0,
\label{prb}
\end{equation}
which prescribes a limitation on the model parameter $A$ which can be further utilize to find the radius of the compact model.\\
Furthermore Eqs.~(\ref{nu2})-(\ref{prb}) generate the mathematical expressions of model parameter as,
\begin{eqnarray}
A &=& \frac{4 M}{R^2 (3 R-4 M)}, \nonumber \\
C &=& {5 \over 2}\sqrt{1-{2 M \over R}},\nonumber \\
D &=& \sqrt{\frac{M}{2 R^3}}. \label{eqacd}
\end{eqnarray}

\begin{figure}[!htbp]
\begin{center}
\begin{tabular}{rl}
\includegraphics[width=9.5cm]{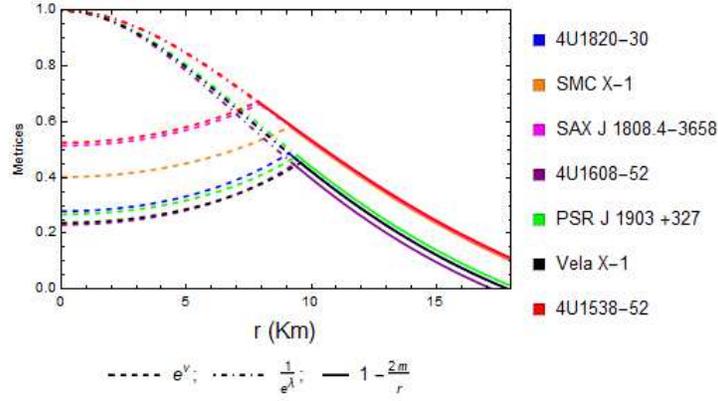}
\\
\end{tabular}
\end{center}
\caption{Junction conditions are satisfied at the boundary for each compact stars.} \label{figmatch}
\end{figure}
In Fig.~\ref{figmatch}, we have illustrated the junction condition of the interior metric potentials with the exterior Schwarzschild metric at the boundary for each stars and it depicts the smooth matching of the boundary conditions.

\begin{table}[tbp]
\centering
\setlength{\tabcolsep}{.3\tabcolsep}
\begin{tabular}{|c|c|c|c|c|c|c|}
\hline
{\bf Pulsar}  & {\bf Mass ($M_{\odot}$)}  & {\bf Radius (Km)} &  ${\bf A }$ & ${\bf C }$ & ${\bf D }$ \\
\hline
$4$U$1820-30$  & $1.58$ & $9.1$ & ~~~$0.00626159$~~~ & ~~~$\pm 1.74607$~~~ & ~~~$\pm 0.0393231$~~~  \\

SMC X-$1$ & $1.29$ & $9.13$ & $0.00461632$ & $\pm 1.90917$ & $\pm 0.0353565$ \\

SAX J $1808.4-3658$  & $0.9 $  & $7.951$ & $0.00452972$ & $\pm 2.04034$ & $\pm 0.0363387$ \\
$4$U$1608-52$  & $1.74$ & $9.3$ & $0.00673108$ & $\pm 1.67344$ & $\pm 0.0399421$ \\
PSR J $1903+327$    & $1.667$  & $9.438$ & $0.00597525$ & $\pm 1.73016$ & $\pm 0.038241$ \\
Vela X-$1$       & $1.77$  & $9.56$ & $0.00626551$ & $\pm 1.68415$ & $\pm 0.0386528$ \\
$4$U$1538-52$     & $0.87$  & $7.866$ & $0.00449277$ & $\pm 2.05201$ & $\pm 0.0363086$ \\
\hline
\end{tabular}
\caption{\label{table1}Values of model parameters for different compact objects.}
\end{table}
\section{\label{sec6}Analysis of the features of the model}
This section contains the inspection of various properties of the interior of the stellar structure. The significant features such as regularity, causality and stability criterion are discussed using numerical calculations and generating graphs. Our spectrum of discussions revolve around the pulsars $4$U$1820-30$ (Mass = $1.58~M\odot$, radius = $9.1$ km \cite{GRDDD}), SMC X-$1$ (Mass = $1.29~M\odot$, radius $= 9.13$ km \cite{RCKSR}), SAX J $1808.4-3658$ (Mass $= 0.9~M\odot$, radius $= 7.951$ km \cite{elebert}), $4$U$1608-52$ (Mass $= 1.74~M\odot$, radius $= 9.3$ km \cite{GOCW}), PSR J$1903+327$ (Mass $= 1.667~M\odot$, radius $= 9.438$ km \cite{freire}), Vela X-$1$ (Mass $= 1.77~M\odot$, radius $= 9.56$ km \cite{rawl}) and $4$U$1538-52$ (Mass $= 0.87~M\odot$, radius $= 7.866$ km \cite{rawl}) .

\subsection{Regularity condition}
\begin{figure}[!htbp]
\centering
\includegraphics[width=.49\textwidth]{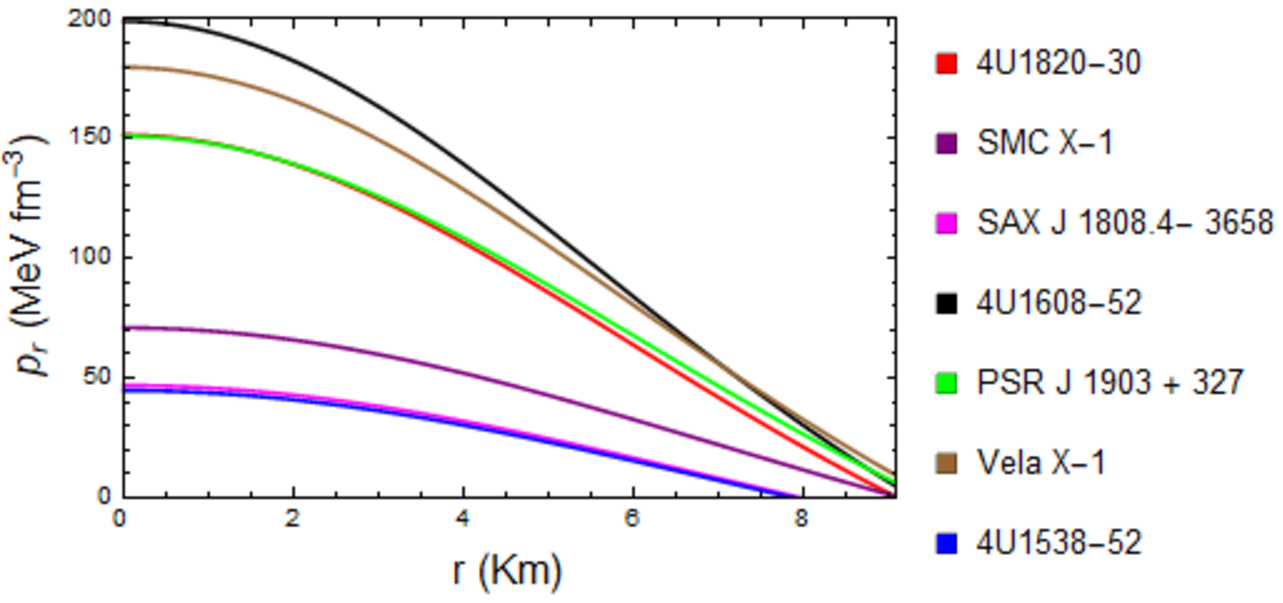}
\hfill
\includegraphics[width=.49\textwidth]{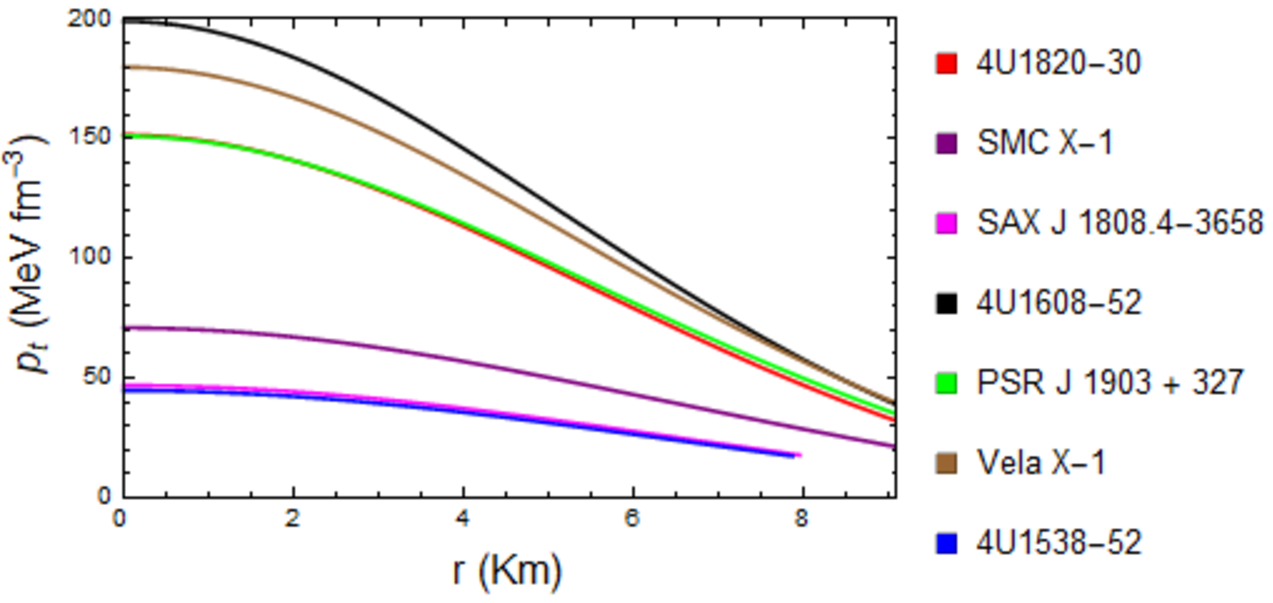}

\caption{\label{figpressure} Radial (left) and Transverse (right) pressures for different compact stars.}
\end{figure}

\begin{figure}[!htbp]
\centering
\includegraphics[width=.49\textwidth]{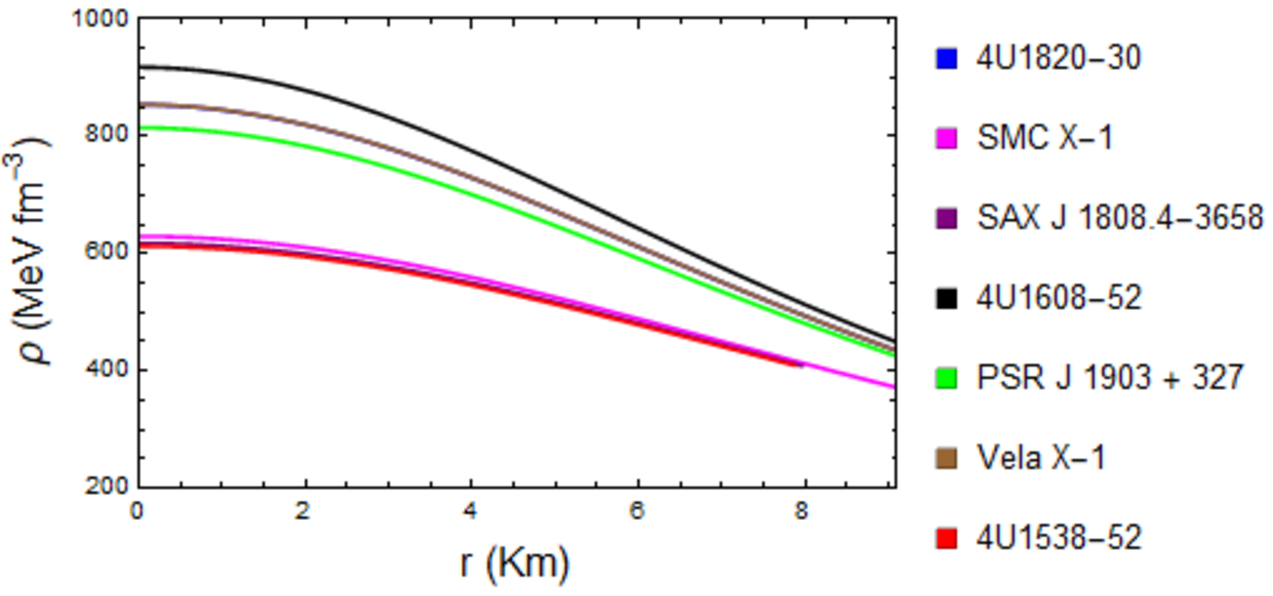}
\hfill
\includegraphics[width=.49\textwidth]{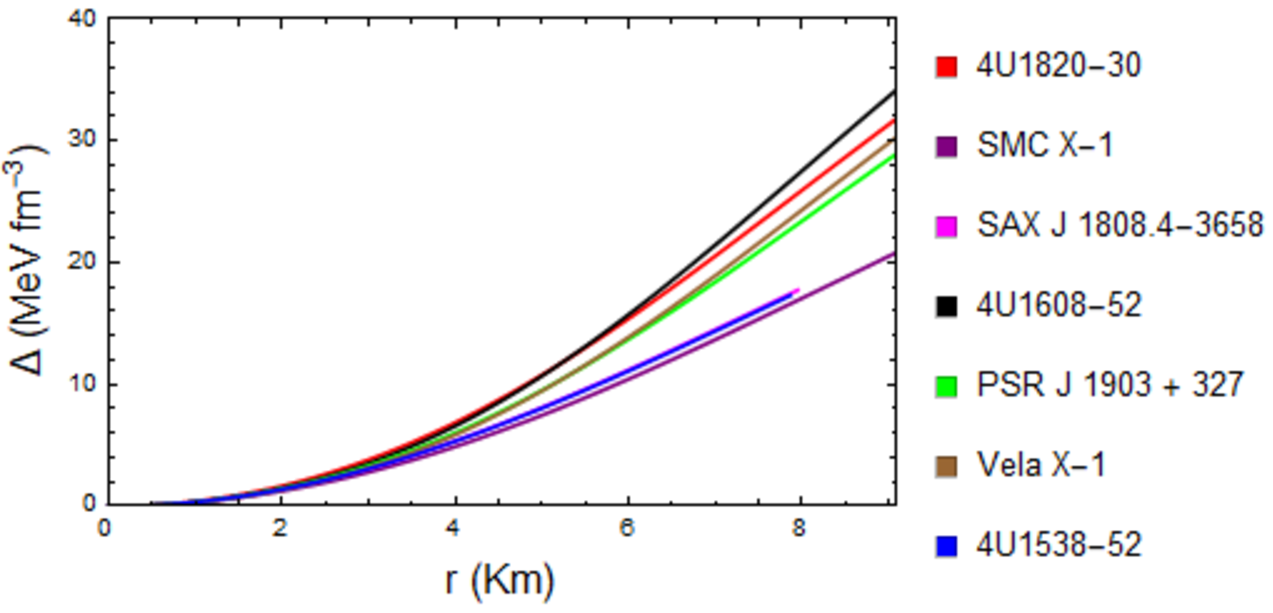}
\caption{\label{figda} Behavior of energy density (left) and anisotropy (right) with respect to the radial coordinate $r$ for various compact stars.}
\end{figure}

To be a physically viable model of an anisotropic compact star, the model ought to satisfy some regularity conditions throughout the interior of the structure \cite{DL,HS,leibovitz,PMP, pant}.\\
(i) The spacetime and hence the solutions need to be free from any singularity i.e. the energy density $\rho$ and pressures $(p_r,~p_t)$ should be finite and positive throughout the star.
Also $e^{\lambda(r)}$ and $e^{\nu(r)}$ should be non zero and finite. Here $e^{\lambda(r)} \big|_{r=0} = 1$ and $e^{\nu(r)} \big|_{r=0} = C^2$ i.e. both are non zero and finite at the center. Also Fig.~\ref{figmp} shows that the metric potentials are positive throughout the stellar structure.\\
(ii) Energy density and pressures must be maximum at the center and monotonically decreasing towards the boundary of the star. Fig.~\ref{figpressure} depicts that both the pressures are monotonically decreasing function of $r$ with a maximum value at the center and radial pressure vanishes at the boundary for each of the stars. Moreover from Fig.~\ref{figda}, it can be seen that energy density is monotonically decreasing in nature and the maximum value can be obtained at the center of the compact model. Analytically, $\frac{d\rho}{dr} \big|_{r =0} = 0$ and $\frac{d^2\rho}{dr^2} \big|_{r =0} < 0$. Also Fig.~\ref{figda} also represents the positive nature of anisotropy inside the stellar interior, which is an important feature for a stable compact model as suggested by Gokhroo and Mehra \cite{GM}.

Central density, central radial and transverse pressures are given as,
 \begin{eqnarray}
 \rho(0) &=& \frac{9 A}{2} ,\\
 p_r(0) = p_t(0)&=&-\frac{A \left(5 \sqrt{3} D - 3 C \sqrt{\frac{A}{2}}\right)}{2 \left( \sqrt{3} D - C \sqrt{\frac{A}{2}} \right)}.
 \end{eqnarray}
Since $A$ is positive so central density is always positive. Also, equality of both the pressures at the center indicates the absence of anisotropy at $r = 0$.

To configure a stable model it is require to satisfy Zeldovich's Condition for pressure and density which states that $\frac{p_r}{\rho}$ must be $\leq 1$ at the center \cite{ZN}. Therefore,
 \begin{eqnarray}
  -\frac{5 \sqrt{3} D - 3 C \sqrt{\frac{A}{2}}}{9 \left( \sqrt{3} D - C \sqrt{\frac{A}{2}} \right)} \leq 1,\\
or, \frac{D}{C} \geq \frac{\sqrt{6 A}}{7}. \label{dc1}
 \end{eqnarray}

Again the gradient of density and pressures can be expressed as,
\begin{eqnarray}
 \frac{d\rho}{dr} &=& -\frac{3 A^2 r (5 + A r^2)}{(1 + A r^2)^3},\\
 \frac{dp_r}{dr} &=&\frac{A^2 \left( \sqrt{\frac{A r^2}{2 - A r^2}} \left(6 A C^2 - 3 A^2 C^2 r + 15 D^2(2 - A r^2) \right)-\sqrt{3} A C D r (17 + 7 A r^2) \right) }{\sqrt{\frac{A }{2 - A r^2}}(2 - A r^2)^2 (1 + A r^2)^2 \left( \sqrt{3} D  - C \sqrt{\frac{A }{2 - A r^2}} \right)^2}, \\
 \frac{dp_t}{dr}&=&\frac{A^2 \left(\sqrt{\frac{A r^2}{2 - A r^2}} \left(12 A C^2 (2 - A r^2)+ 6 D^2 (2 - A r^2)(9 + A r^2) \right) \right)}{2 \sqrt{\frac{A}{2 - A r^2}}(2 - A r^2)^2 (1 + A r^2)^3 \left(\sqrt{3}D  - C \sqrt{\frac{A}{2 - A r^2}}\right)^2} \nonumber \\
&+&\frac{\sqrt{3} A^3 C D r (A^2 r^4 + 23 A r^2 -62)}{2 \sqrt{\frac{A }{2 - A r^2}}(2 - A r^2)^2 (1 + A r^2)^3 \left(\sqrt{3}D  - C \sqrt{\frac{A }{2 - A r^2}}\right)^2}.
\end{eqnarray}

\begin{figure}[!htbp]
\begin{center}
\begin{tabular}{rl}
\includegraphics[width=9.5cm]{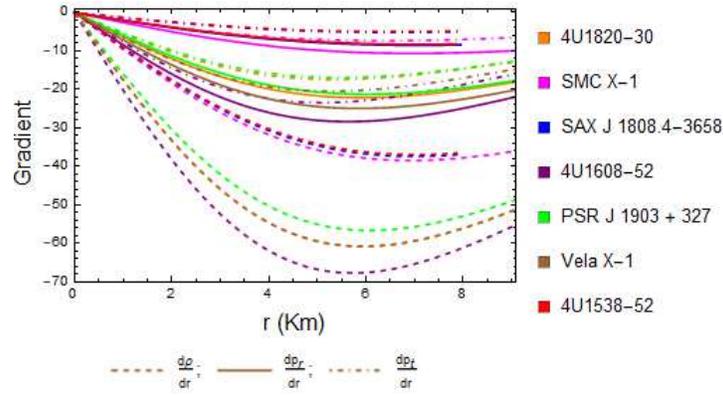}
\\
\end{tabular}
\end{center}
\caption{Energy density gradient, pressures gradient for different compact stars.} \label{figgrad}
\end{figure}

For a stable compact model $\frac{dp_r}{dr} \big|_{r =0} = \frac{dp_t}{dr} \big|_{r =0} = 0$  and  $\frac{d^2 p_r}{dr^2} \big|_{r =0} < 0$ and $\frac{d^2 p_t}{dr} \big|_{r =0} < 0$ need to satisfy inside the star  i.e. gradient of density and pressures are negative within $0 < ~r< ~R$. The negative nature of the gradient of density and pressures are shown graphically in Fig.~\ref{figgrad}.

Also $p_r(0) = ~p_t(0) \geq 0$ gives the relation
\begin{equation}
\frac{\sqrt{3A}}{5\sqrt{2}} \leq \frac{D}{C} \leq \sqrt{\frac{A}{6}}.
\label{dc2}
\end{equation}
Combining Eqs.~(\ref{dc1}) and (\ref{dc2}) we get the bounds on the model parameters as
\begin{equation}
\frac{\sqrt{6A}}{7} \leq \frac{D}{C} \leq \sqrt{\frac{A}{6}}. \label{dc}
\end{equation}
\subsection{Kretschmann Scalar}
\begin{figure}[!htbp]
\begin{center}
\begin{tabular}{rl}
\includegraphics[width=9.5cm]{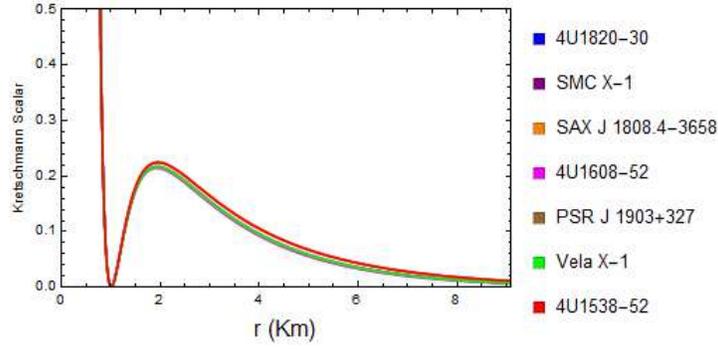}
\\
\end{tabular}
\end{center}
\caption{Kretschmann Scalar for different compact stars.} \label{figks}
\end{figure}

In General Relativity, scalars are used to look for any singularity present in the metric. The simplest is the Ricci Scalar but since for vacuum solution, Ricci scalar is zero everywhere so to find any physical singularity present in the spacetime Kretschmann Scalar is used. The approach is quite straightforward. For any line element,
\begin{equation}
ds^2= e^{2\mu}dt^2 - e^{2\kappa}dr^2 -
e^{2\zeta} (d\theta^2 + \sin^2 \theta d\phi^2),
\end{equation}
Kretschmann Scalar is calculated as,
\begin{equation}
K = 4 K_1^2 + 8K_2^2 + 8 K_3^2 + 4 K_4^2,
\end{equation}
where,
\begin{eqnarray}
K_1 &=& e^{-(\mu +\kappa)}\frac{d}{dr}\left(\frac{d\mu}{dr}e^{\mu-\kappa }\right), \nonumber \\ \nonumber
K_2 &=& e^{-2\kappa}\frac{d\zeta}{dr}\frac{d\mu}{dr}, \\ \nonumber
K_3 &=& e^{-(\kappa+\zeta)} \frac{d}{dr} \left(e^{\zeta-\kappa} \frac{d\zeta}{dr}\right), \\ \nonumber
K_4 &=& -e^{-2\zeta}+e^{-2\kappa}\left(\frac{d\zeta}{dr}\right). \nonumber
\end{eqnarray}
Computing all the components we have obtained Kretschmann Scalar for various compact objects and it is shown graphically in Fig.~\ref{figks}.
The divergence of Kretschmann Scalar at $r=0$ depicts that there is no singularity at the center of the compact model, although there is a singularity at $r=1$.
\subsection{The Tolman-Oppenheimer-Volkoff or TOV Equation}
To examine the stability of the model it is important to examine the equilibrium condition of the model using TOV equation. This stability equation given by Tolman \cite{tolman} and Oppenheimer and Volkoff \cite{OV} symbolizes the internal structure of a spherically symmetric static compact object which is in equilibrium in presence of anisotropy. The generalized TOV equation can be expressed as \cite{varela,ponce},
\begin{equation}
-{1 \over r}\left(\rho(r)+p_r(r)\right) {d\nu \over dr}-{dp_r(r) \over dr}+{2(p_r-p_t) \over r}=0.\nonumber
\end{equation}
It can also be written as,
\begin{eqnarray}
-{M_g\left(\rho(r)+p_r(r)\right) \over r^2}~e^{(\lambda-\nu)/2}-{dp_r(r) \over dr}+{2\Delta(r) \over r}=0,\nonumber
\\\label{tov1}
\end{eqnarray}
where $M_g(r)$ is the effective gravitational mass inside a sphere of radius `$r$' and it can be derived using Tolman-Whittaker mass formula is given by,
\begin{equation}
M_g(r)={1 \over 2}r^2 e^{\frac{\nu-\lambda}{2}} {d\nu \over dr}.
\label{tov2}
\end{equation}

The TOV equation can be expressed in a simple form to describe the equilibrium condition by defining the forces as gravitational forces($F_g$), hydrostatic forces($F_h$) and anisotropic forces($F_a$).
Thus,
\begin{equation}
F_g(r) + F_h(r) + F_a(r) =0,
\label{force1}
\end{equation}
where,
\begin{eqnarray}
\text{gravitational force},~ F_g(r)& =& -\frac{\nu'\left(\rho(r)+p_r(r)\right)}{2},\nonumber \\
\text{hydrostatic force},~ F_h(r)& =& -{dp_r(r) \over dr},\nonumber \\
\text{anisotropic force},~ F_a(r) &=& {2\Delta(r) \over r}.\label{force2}
\end{eqnarray}
For the prescribed model the expressions for these forces become,
\begin{eqnarray}
F_g(r) &=& \frac{A^3 D r \left( 3 C \sqrt{\frac{A}{2 -A r^2}} - \sqrt{3}D (2 - A r^2) \right)}{\sqrt{\frac{A (2 - A r^2)}{3}} (1 + A r^2)^2 \left(\sqrt{3}D - C \sqrt{\frac{A}{2-A r^2}}\right)\left( A C - \sqrt{3} D \sqrt{A(2-A r^2)}\right)}, \nonumber \\
F_h(r) &=& \frac{A^2 r \left(\sqrt{3}A C D (17-7A r^2)- 3 A C^2 \sqrt{\frac{A}{2-A r^2}} (2 - A r^2) - 15 D^2 \sqrt{A}(2-A r^2)^{3 \over 2} \right)}{\sqrt{A} (2 -A r^2)^{3 \over 2} (1+ A r^2)^2 \left(\sqrt{3}D- C\sqrt{\frac{A}{2-A r^2}}\right)^2}, \nonumber \\
F_a(r) &=& \frac{A^2 r (4\sqrt{3} D - 3 C \sqrt{\frac{A}{2-A r^2}})}{(1+A r^2)^2(\sqrt{3}D - C \sqrt{\frac{A}{2 - A r^2}})}. \label{force3}
\end{eqnarray}
\begin{figure}[!htbp]
\begin{center}
\begin{tabular}{rl}
\includegraphics[width=9.5cm]{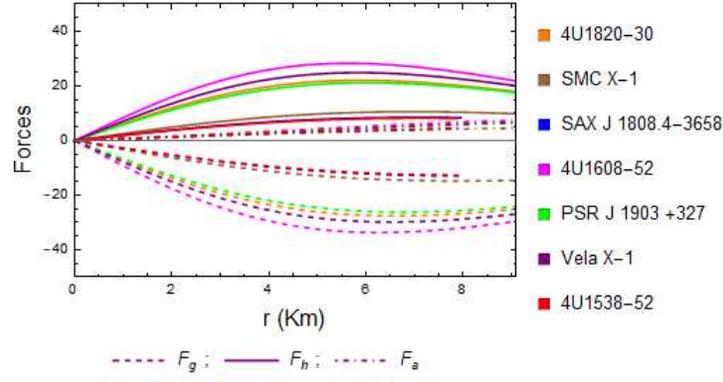}
\\
\end{tabular}
\end{center}
\caption{Different forces for different compact stars.} \label{figforce}
\end{figure}
Equations alluded in Eq.~(\ref{force3}) are examined graphically in the Fig.~\ref{figforce}. It can clearly be seen that negative gravitational force is balanced by the amalgamation of hydrostatic and anisotropic forces to keep the model in equilibrium.
\subsection{Energy condition}
\begin{figure}[!htbp]
\begin{center}
\begin{tabular}{rl}
\includegraphics[width=9.5cm]{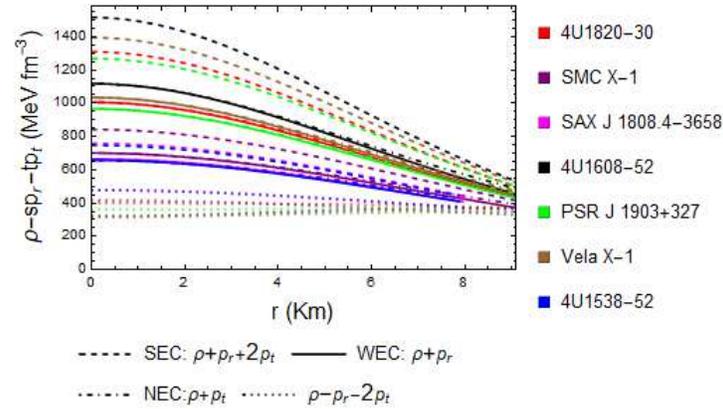}
\\
\end{tabular}
\end{center}
\caption{Behavior of different energy conditions on different compact stars. The nature of $\rho - p_r - 2 p_t$ is also plotted.}\label{figenergy}
\end{figure}

The acceptability of our model depends on fulfillment of some energy conditions namely,  Null energy condition (NEC), Weak energy condition (WEC), Strong energy condition (SEC) and Dominant Energy Condition (DEC). All these energy conditions are  some inequalities corresponding to stress-energy tensor and are defined as,
\begin{eqnarray}
NEC_r  &:&  \rho(r) + p_r(r)\geq 0,~~~~ NEC_t  :  \rho(r) + p_t(r)\geq 0.\nonumber
\\
WEC_r  &:& \rho(r)\geq 0,~~ \rho(r) + p_r(r)\geq 0,~~~
WEC_t  : \rho(r)\geq 0,~~ \rho(r) + p_t(r)\geq 0.\nonumber
\\
SEC  &:& \rho(r) + p_r(r)+2p_t(r)\geq 0. \nonumber \\
DEC &:& \rho(r) \geq p_r(r),~ p_t(r).
\label{energy}
\end{eqnarray}
Analytically NEC suggests that an eyewitness crossing a null bend will quantify the surrounding energy density to be non-negative. WEC implies that energy density estimated by an eyewitness traversing a time like bend is positive. SEC implies that the trace of tidal tensor estimated by the eyewitness is consistently positive \cite{MESTD}. DEC essentially indicates that to any observer the local energy density appears non-negative and local energy flow vector is non-spacelike \cite{HE}.\\
We have discussed the energy conditions on our model graphically in Fig.~\ref{figenergy}, by plotting LHS (left hand side) of above inequalities.

Also for distinguishing configuration we can develop some limitation on our model parameter from inequalities in Eq.~(\ref{energy}). Specifically we obtain the following relations for the center ($r = 0$) of the stellar structure,
\begin{eqnarray}
NEC_r  &:&  \rho(0) + p_r(0)\geq 0,~~~~ NEC_t  :  \rho(0) + p_t(0)\geq 0.\nonumber
\\
WEC_r  &:& \rho(0)\geq 0,~~ \rho(0) + p_r(0)\geq 0.\nonumber
\\
WEC_t  &:& \rho(0)\geq 0,~~ \rho(0) + p_t(0)\geq 0.\nonumber
\\
SEC  &:& \rho(0) + p_r(0) + 2p_t(0)\geq 0.\nonumber \\
DEC &:& \rho(0)\geq p_r(0),~p_t(0) \nonumber \\
&or& \frac{9A}{2} \geq \frac{A\left(5\sqrt{3}D- C \sqrt{A \over 2}\right)}{2\left(\sqrt{3}D - C \sqrt{A \over 2}  \right) } \nonumber \\
&or& {D \over C} \geq \sqrt{2 A \over 3}.
\label{energy2}
\end{eqnarray}
\subsection{Causality condition}
\begin{figure}[!htbp]
\centering
\includegraphics[width=.49\textwidth]{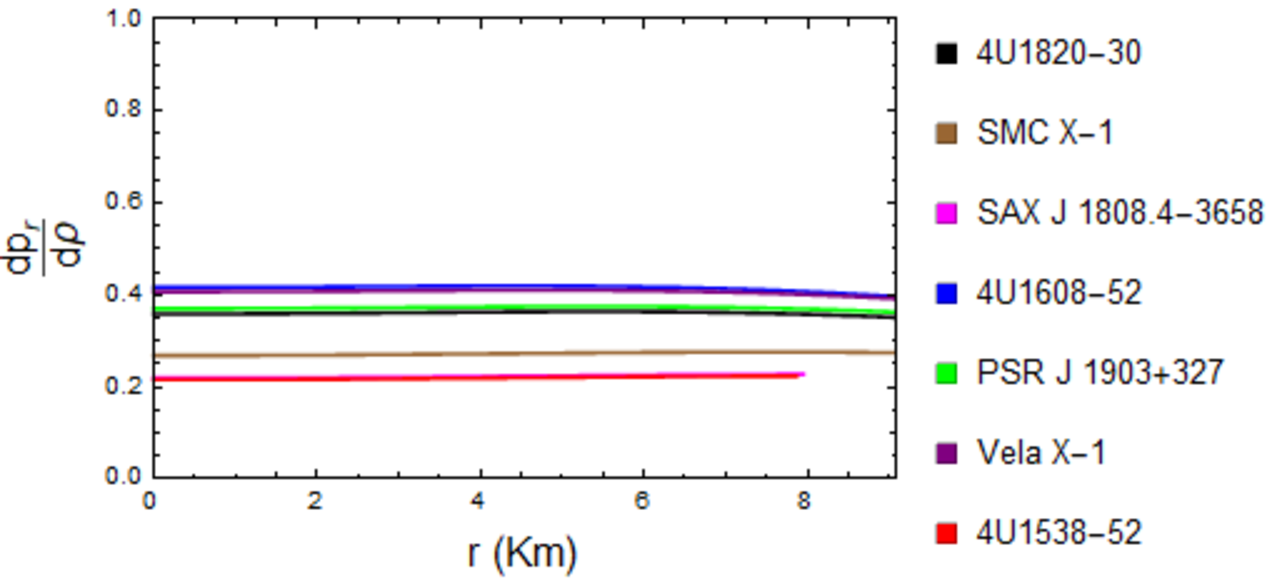}
\hfill
\includegraphics[width=.49\textwidth]{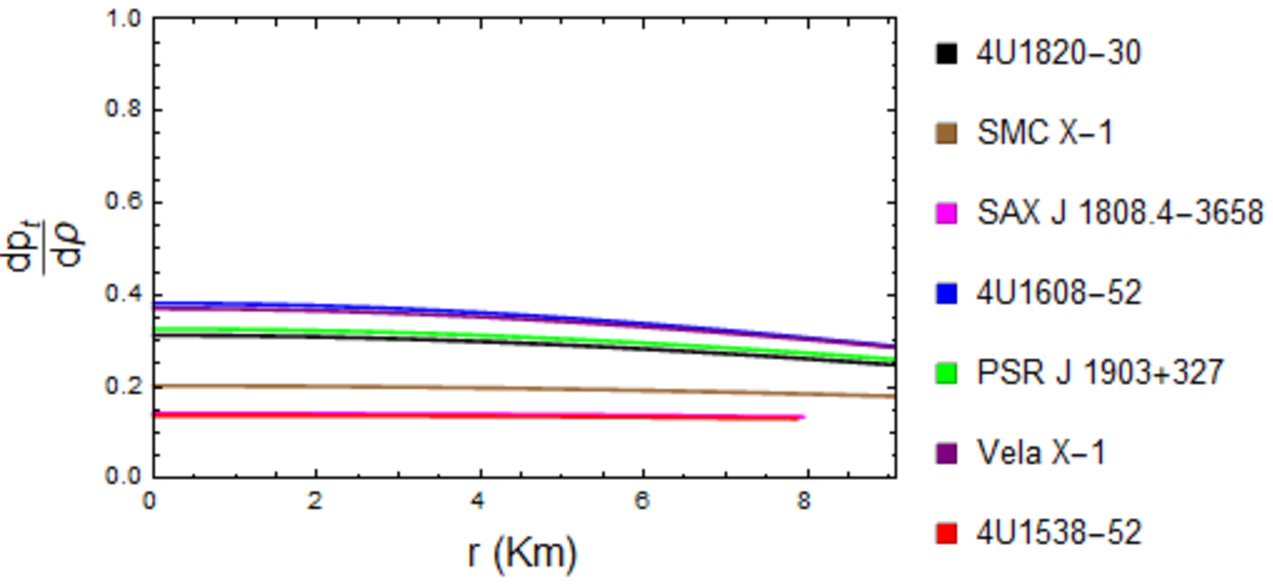}
\caption{\label{figvsrvst} Radial (left) and Transverse (right) sound speeds for different compact stars.}
\end{figure}

To study the stability of an anisotropic fluid stellar,  L. Herrera \cite{herrera92} proposed the  $cracking$ $method$ or overturning method which states that the velocity of sound speeds (radial and transverse) should never exceed the speed of light inside the star i.e. $v^2 = {dp \over d\rho} < 1$ should be maintained inside the stellar, taking the velocity of light $c = 1$. Also Le Chatelier's principle allows the matter of the star to satisfy $\frac{dp}{d\rho} \geq 0$ to be a stable configuration \cite{NKG}. The sound velocity inside the compact star is expressed by,
\begin{equation}
v_r(r)=\sqrt{{dp_r(r) \over d\rho(r)}},~~~v_t(r)=\sqrt{{dp_t(r) \over d\rho(r)}}.
\end{equation}
Combining the above conditions the causality condition becomes $0 \leq v_r(r),~v_t(r) < 1$. Fig.~\ref{figvsrvst} supports the fulfillment of causality condition of the prescribed model.
 Now using the concept of cracking, Abreu et al. \cite{abreu} provided the stability conditions with respect to the stability factor $\left(= \{v_t(r)\}^2 - \{v_r(r)\}^2\right)$ for anisotropic fluid model. The conditions are {\bf:} \textbf{(i)} The region is potentially stable if $ -1 < \{v_t(r)\}^2 - \{v_r(r)\}^2 \leq 0$ and \textbf{(ii)} The region is potentially unstable if $ 0 < \{v_t(r)\}^2 - \{v_r(r)\}^2 < 1$.\\

\begin{figure}[!htbp]
\centering
\includegraphics[width=.49\textwidth]{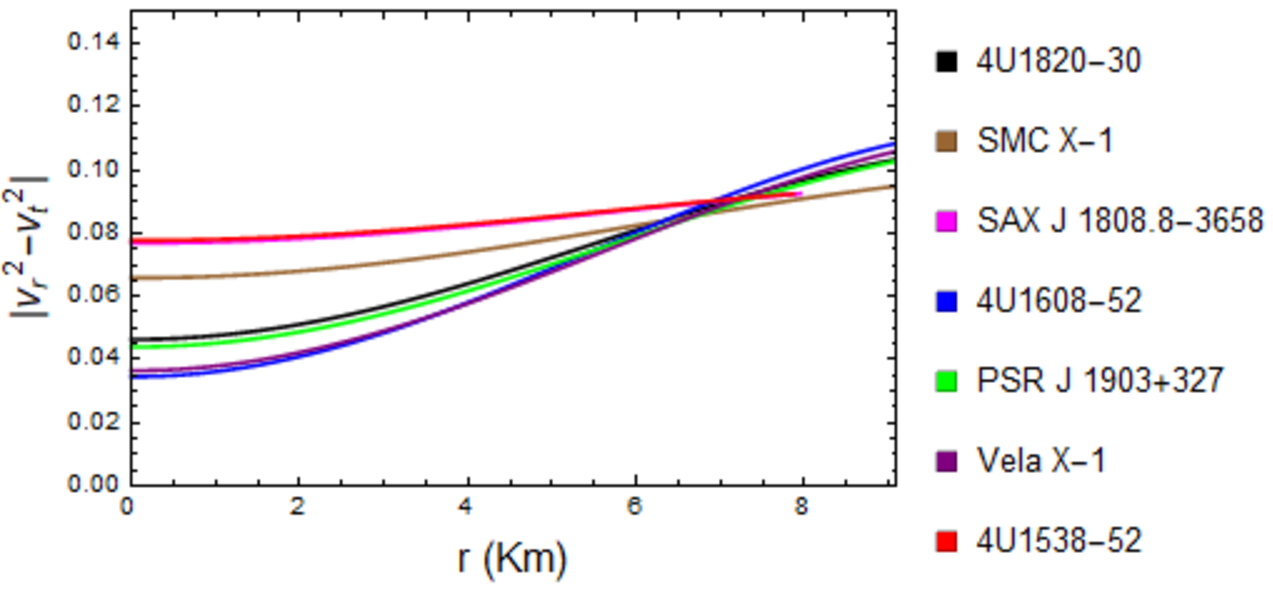}
\hfill
\includegraphics[width=.49\textwidth]{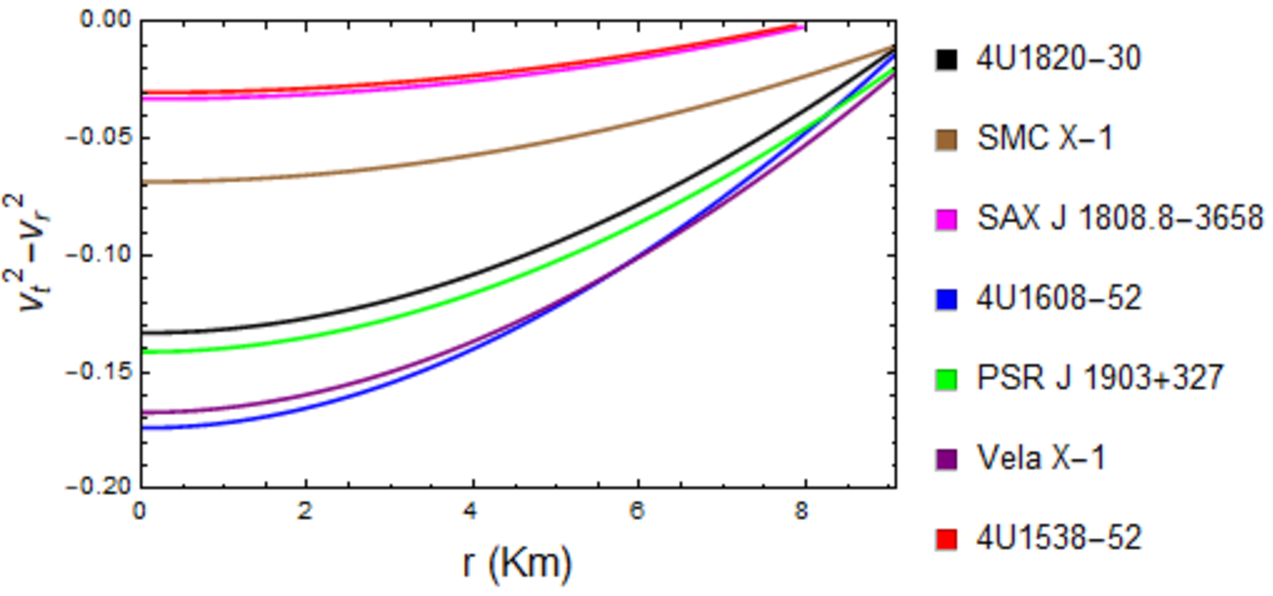}
\caption{\label{figsound}  Variation of absolute difference (left) and }variation of difference of sound speeds (right) with radial coordinate for different compact stars.
\end{figure}

The expressions for velocity of sound speeds are given below,
\begin{eqnarray}
v_{r}^2 &=& - \frac{\sqrt{\frac{A}{2 - A r^2}}(2 - A r^2) (1 + A r^2)(3 A C^2 + 30 D^2 - 15 A D^2 r^2) }{3 \sqrt{\frac{A}{2 - A r^2}} (2 - A r^2)^2 (5 + A r^2) \left(\sqrt{3}D - C \sqrt{\frac{A}{2 - A r^2}}\right)^2} \nonumber \\
&+& \frac{\sqrt{3} A C D (1 + A r^2)(17 - 7 A r^2)  }{3 \sqrt{\frac{A}{2 - A r^2}} (2 - A r^2)^2 (5 + A r^2) (\sqrt{3}D - C \sqrt{\frac{A}{2 - A r^2}})^2} , \label{eqvsr} \\
v_{t}^2 &=& - \frac{\sqrt{\frac{A}{2 - A r^2}}\left( 12 A C^2 (2 - A r^2) + 6 D^2 (2 - A r^2)^2 (9 + A r^2) \right) }{6 \sqrt{\frac{A}{2 - A r^2}} (2 - A r^2)^2 (5 + A r^2) (\sqrt{3} D - C \sqrt{\frac{A}{2 - A r^2}})^2} \nonumber \\
&-& \frac{\sqrt{3} A C D (A^2 r^4 + 23 A r^2 -62)}{6 \sqrt{\frac{A}{2 - A r^2}} (2 - A r^2)^2(5 + A r^2) (\sqrt{3} D - C \sqrt{\frac{A}{2 - A r^2}})^2}. \label{eqvst}
\end{eqnarray}
 Clearly the conditions $ |{v_r}^2-{v_t}^2| <~1$  and $ -1 < \{v_t(r)\}^2 - \{v_r(r)\}^2 \leq 0$ are satisfied as depicted in Fig.~\ref{figsound}.

\subsection{Adiabatic index }
Stability of anisotropic compact star depends on the adiabatic index which is essentially the ratio of specific heat at constant pressure to the specific heat at constant volume. The adiabatic index determines the stability and the stiffness of the equation of state and it is defined as,
\begin{eqnarray}
\Gamma(r) = {\rho(r)+p(r) \over p(r)}~{dp(r) \over d\rho(r)}.
\end{eqnarray}
 For the Newtonian limit, any stellar configuration will maintain its stability if adiabatic gravitational collapse $\Gamma (r) > 4/3$ \cite{HH} and stellar structure will become catastrophic if $< 4/3$ \cite{bondi}. Also, Chan et al.\cite{chan} have suggested that this condition changes depending on the nature of anisotropy for a relativistic fluid sphere. Additionally, Knutsen \cite{knutsen} showed that adiabatic index $\Gamma$ is more than $1$ if the ratio of density and pressure is monotonically decreasing outwards.\\

\begin{figure}[!htbp]
\begin{center}
\begin{tabular}{rl}
\includegraphics[width=9.5cm]{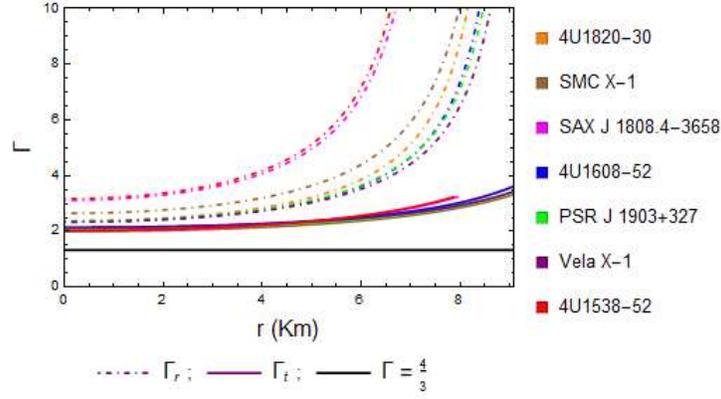}
\\
\end{tabular}
\end{center}
\caption{Adiabatic indices for different compact stars.} \label{figai}
\end{figure}

 For our solution, the value of adiabatic indices $\Gamma_r(r)$ and $\Gamma_t(r)$ are more than 4/3 throughout the outer region of a compact star, as evident from Fig.~\ref{figai}.

\subsection{Harrison-Zeldovich-Novikov criterion}
\begin{figure}[!htbp]
\centering
\includegraphics[width=.49\textwidth]{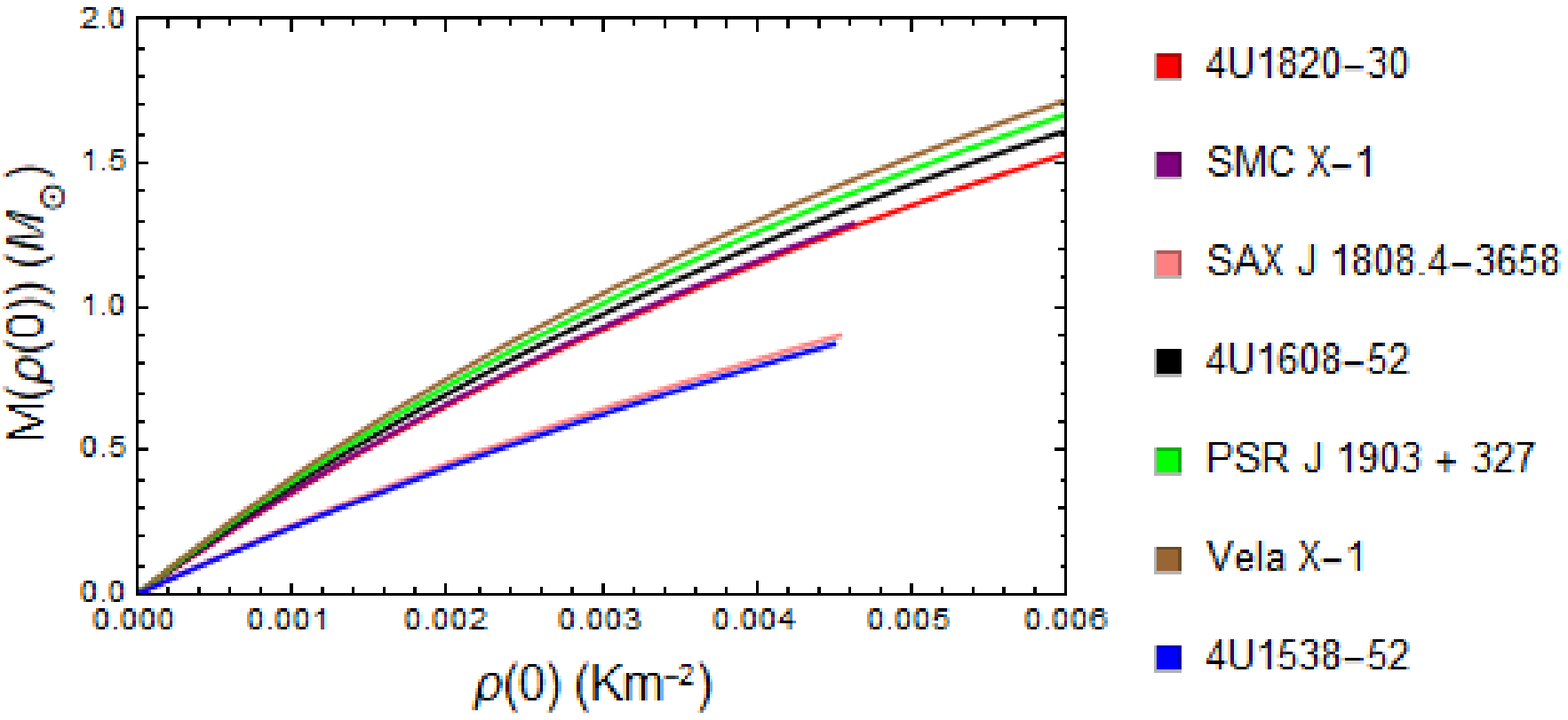}
\hfill
\includegraphics[width=.54\textwidth]{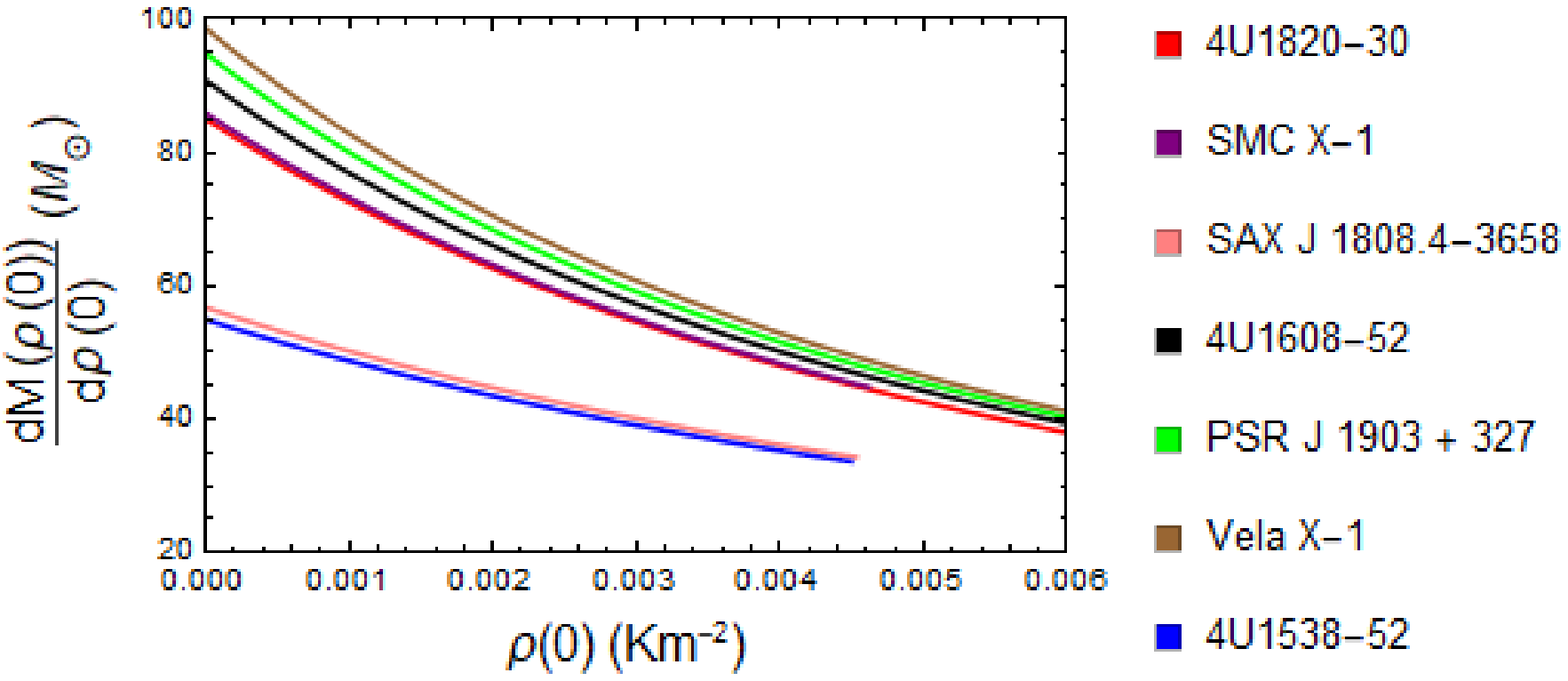}
\caption{\label{fighzn} Variation of $M$ and ${dM \over d\rho(0)}$ with respect to the central density $\rho(0)$ for different compact stars.}
\end{figure}

Since compact models that are only in stable equilibrium, are of astrophysical interest so any acceptable model should satisfy the static stability criterion. The stability condition for a compact star with respect to `$r$' requires the calculation of eigen-frequency of the fundamental mode \cite{HPY} of radial pulsation without any nodes as described by Chandrasekhar \cite{chandrasekhar}. The complexity of stability criterion was later simplified by Harrison-Zeldovich-Novikov \cite{harrison,ZN}. As per their suggestions, for stability of a compact object the mass should also increase with the increase of the central density $\rho(0)$ i.e. $\frac{d M(\rho(0))}{d \rho(0)}~>~0$ to be a stable structure. The Harrison-Zeldovich-Noikov criterion is a necessary condition, it is not an sufficient one. We write the mass function as function of central density as following,
\begin{eqnarray}
M(\rho(0))&=& \frac{3 R^3 \rho(0)}{2 \left(2 \rho(0) R^2 + 9\right)}, \nonumber \\
\frac{d M(\rho(0))}{d \rho(0)} &=& \frac{27 R^3}{2 \left( 2 \rho(0) R^2 + 9\right)^2}. \label{eqhzn}
\end{eqnarray}
Clearly the fulfillment of Harrison-Zeldovich-Novikov conditions are shown graphically for several stars as shown in Fig.~\ref{fighzn}.

\subsection{Buchdahl Limit}
\begin{figure}[!htbp]
\begin{center}
\begin{tabular}{rl}
\includegraphics[width=8.8cm]{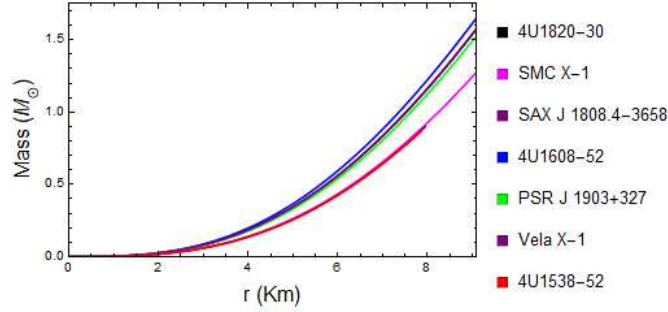}
\end{tabular}
\end{center}
\caption{Mass function of different compact star. Here mass function is shown to be monotonically increasing function of $r$. }\label{figmass}
\end{figure}

The mass function of the proposed model is defined in Eq.~(\ref{massf}) as $m(r)~=~\frac{3 A r^3}{4 (1 +A r^2)}$ which is directly proportional to $r$ i.e. $\lim_{r \to 0} m(r) = 0$ implying the regularity of the mass function at the center of the star. Fig.~\ref{figmass} graphically depicts the mass functions of various compact objects.

For spherically symmetric configuration, the ratio of mass to the radius of a compact object is supposed to fall within the limit ${2M \over R}~<~{8 \over 9}$ (considering $8\pi G~=~c~=~1$) as suggested by Buchdahl \cite{buchdahl59}. This condition, named after Buchdahl, is clearly satisfied for our model as shown in Table~\ref{table2}.
\subsection{Mass-radius relationship}
\begin{figure}[!htbp]
\begin{center}
\begin{tabular}{rl}
\includegraphics[width=7cm]{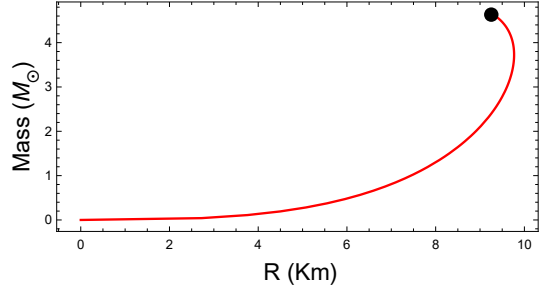}
\\
\end{tabular}
\end{center}
\caption{$(M-R)$ plot for the surface density $\rho(R)~=~9.5 \times 10^{14}~gm~cm^{-3}$. The solid circle denotes the maximum allowable mass of the model.}\label{figmr}
\end{figure}

For dynamical stability opposing gravitational collapse into a black hole, the maximum mass of any model needs to be considered to separate compact star and black holes. In fact, any observed compact objects can be identified as black holes if the maximum mass of the compact object exceeds the allowable maximum mass for a stable compact model \cite{ST,HBPP}. To study the mass-radius relation and to calculate maximum mass we have plotted the $(M-R)$ graph in Fig.~\ref{figmr} considering the surface density $\rho(R)~=~9.5 \times 10^{14}~gm~cm^{-3}$. We have considered surface density which is roughly similar to that chosen by some researchers \cite{SM,ThM,RPSD} to study the mass-radius relationship of a compact model.\\
Fig.~\ref{figmr} depicts the ($M-R$) graph for the prescribed model. The maximum mass for the model is calculated to be $4.632~ M_\odot$ with the radius $9.254$ km. Though our model exceeds the Rhoades and Ruffini limit ($\approx 3.2~ M_\odot$) \cite{RR} of maximum mass for a neutron star, it remains within the prescribed range for the spheres in general relativity with uniform density ($5.2~ M_\odot$) as suggested by Shapiro and Teukolsky \cite{ST}.\\
 We have also predicted masses and radii for several compact stars as given in Table~\ref{table2}. We have computed the masses and radii from EoSs considering EoS parameter $\omega = 0.003$ and it can clearly be seen that predicted masses and radii are almost similar to that of observed values.
\begin{table}[!htp]
\centering
\setlength{\tabcolsep}{.1\tabcolsep}
\begin{tabular}{|c|c|c|c|c|c|c|}
\hline
 Pulsar Name & Observed &  Observed  &  Predicted  &  Predicted  &  Compactness \\
 & Mass ($M_{\odot}$) & Radius (Km) & Mass ($M_{\odot}$) & Radius (Km) & Factor \\ \hline
$4$U$1820-30$  & $1.58$ & $9.1$ & $1.5508$ & $9.0272$ & $0.1717$  \\

SMC X-$1$ & $1.29$ & $9.13$ & $1.2513$ & $8.9873$ & $0.1392$ \\

SAX J $1808.4-3658$  & $0.9 $  & $7.951$ & $0.8460$ & $8.9168$ & $0.0949$ \\

$4$U$1608-52$  & $1.74$ & $9.3$ & $1.7242$ & $7.9172$ & $0.2178$ \\

PSR J $1903+327$    & $1.667$  & $9.438$ & $1.6387$ & $9.2310$ & $0.1775$ \\

Vela X-$1$       & $1.77$  & $9.56$ & $1.7435$ & $9.3755$ & $0.1859$ \\

$4$U$1538-52$     & $0.87$  & $7.866$ & $0.8273$ & $7.7138$ & $0.1072$ \\ \hline
\end{tabular}
\caption{\label{table2}Observed and predicted  mass, radius and compactness factor for different compact objects.}
\end{table}
\subsection{Compactness and surface redshifts}
The dimensionless ratio $\frac{m(r)}{r}$ is known as the compactification factor $u(r)$ of a compact star. The expressions for compactness and surface redshifts are given as following,
\begin{eqnarray}
u(r) &=& \frac{m(r)}{r}= \frac{3 A r^2}{4(1 + A r^2)}, \nonumber \\
z(r) &=& \frac{1 - \left( 1 - 2 u \right)^{ 1 \over 2}}{\left(1 - 2 u \right)^{1 \over 2}} = \sqrt{\frac{2(1 + A r^2)}{2 - A r^2}}-1. \label{eqred}
\end{eqnarray}

\begin{figure}[!htbp]
\centering 
\includegraphics[width=.49\textwidth]{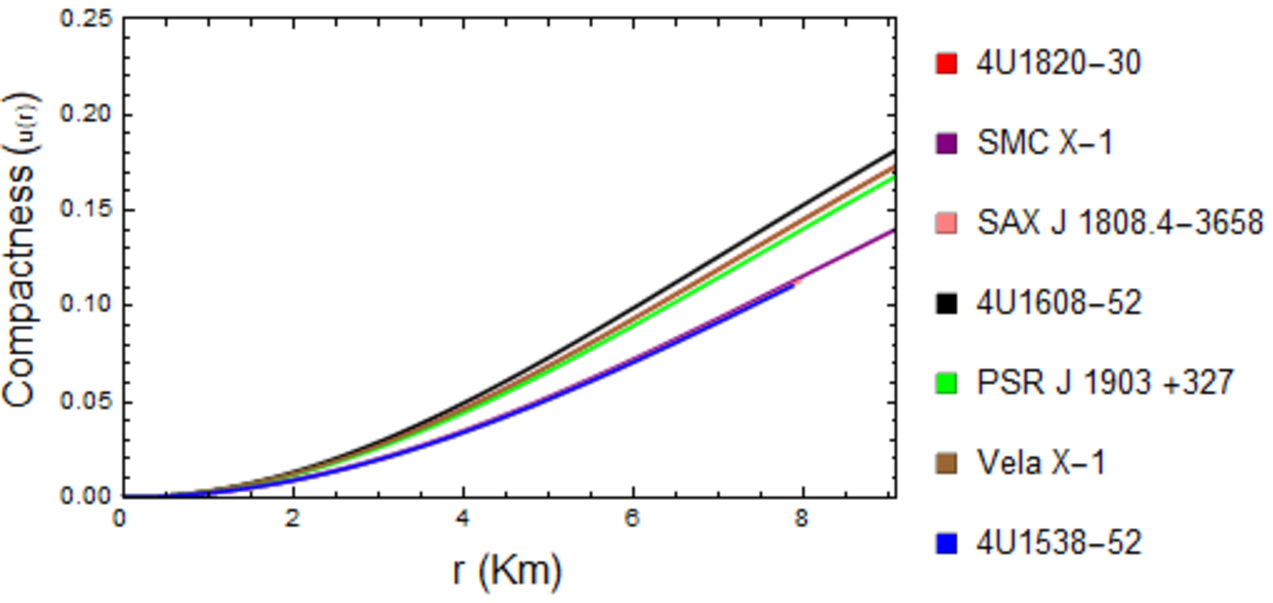}
\hfill
\includegraphics[width=.49\textwidth]{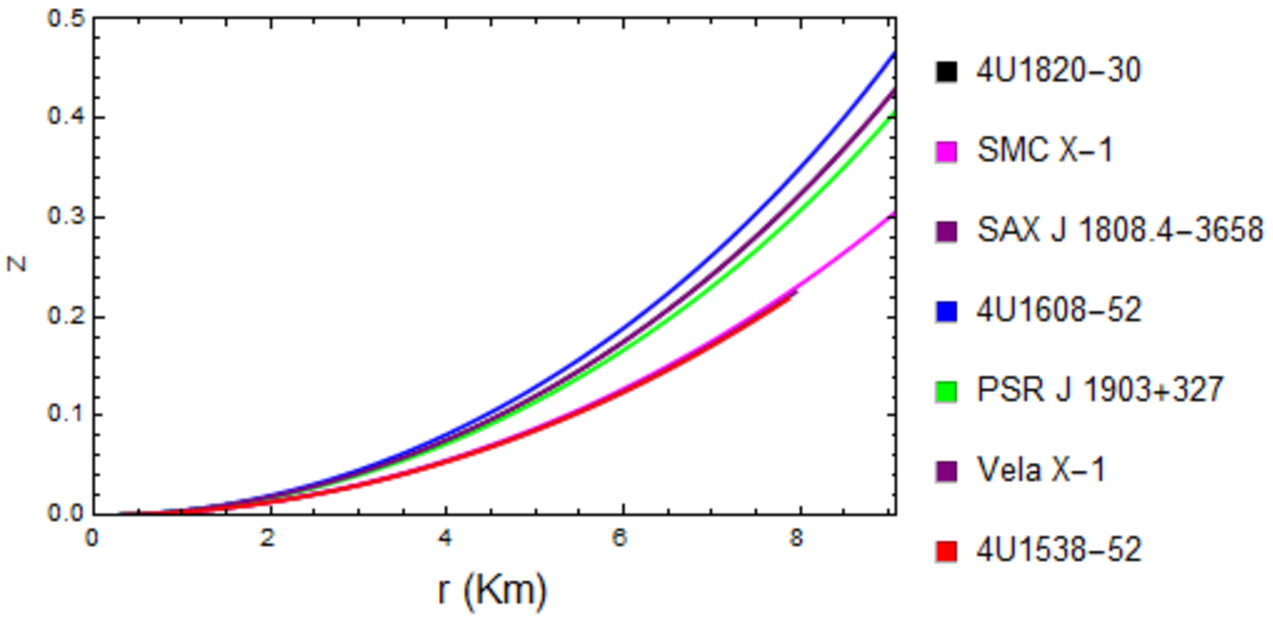}
\caption{\label{figcompred} Compactness (left) and surface redshift (right) with the radial coordinate $r$ for different compact stars.}
\end{figure}

The compactness for our model is depicted in Fig.~\ref{figcompred} which indicates the increasing nature of the compactification factor with respect to radial coordinate `$r$'. From Table.~\ref{table2} it can be seen that the model allows compactness within the range $({1 \over 4}, {1 \over 2})$ as prescribed in  \cite{JT}.

The surface redshifts $z(r)$ is depicted graphically in Fig.~\ref{figcompred} which shows the increasing nature of surface redshifts with the radial coordinate. Studies by several authors have allowed us to specify an upper bound on the surface redshifts. In absence of cosmological constant, $z(r) \leq 2$ holds for an isotropic star as proposed by Barraco and Hamity \cite{BHamity}. Whereas the presence of cosmological constant pushes the upper bound for an anisotropic star a bit higher, $z(r) \le 5$ \cite{bohmer}. Though the maximum acceptable limit for the surface redshift of a compact star is $5.211$ \cite{ivanov}, the model presented in this paper satisfies the range $z(r) \leq 1$ as predicted by Hewish et. al \cite{HBPSC} as can be seen in Table~\ref{table3}.
\subsection{Equation of State}
One of the significant features of a compact star is the description of its equation of state (EoS) i.e. the relation of the pressure with the energy density for barotropic EoS which then eventually designs the mass-radius relation. The form of the barotropic equation of state can be linear, quadratic, polytropic or some other dependence. Clearly different EoSs lead to different $(M-R)$ relations. Several authors have suggested that EoS can be estimated in the form of $p~=~p(\rho)$ i.e. pressure $p$ can be written as the linear function of energy density $\rho$ in the presence of higher density to elucidate the structural properties of a compact object \cite{FO,HZ,PBP,LPMY,DBDRS,GBZGRDD,zdunik,HC,MMKP,MBJKPP}.
We have obtained the exact similar observation as that of \cite{GBZGRDD,MMKP,MBJKPP}.

\begin{figure}[!htbp]
\begin{center}
\begin{tabular}{rl}
\includegraphics[width=9.4cm]{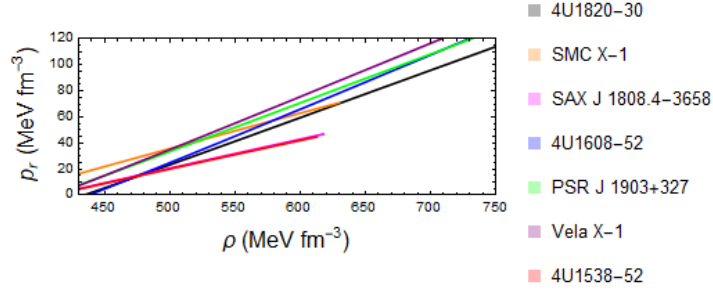}
\\
\end{tabular}
\end{center}
\caption{ Variation of radial pressure with density for different compact star corresponding to the numerical value of constants given in Table~\ref{table1}.}\label{figeos}
\end{figure}

\begin{figure}[!htbp]
\begin{center}
\begin{tabular}{rl}
\includegraphics[width=9cm]{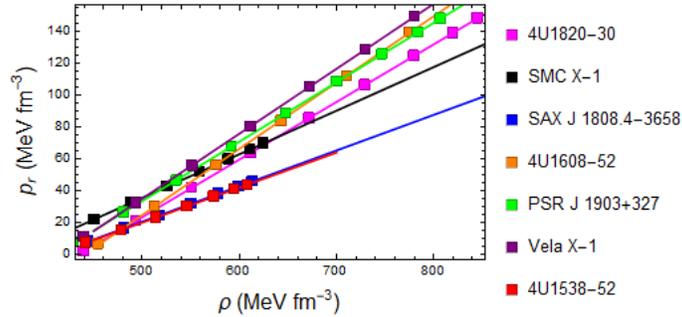}
\\
\end{tabular}
\end{center}
\caption{Best fit for EoSs for each compact star.}\label{figfit}
\end{figure}

For stable configuration the equation of state parameter defined as $\omega_r(r) = \frac{p_r}{\rho}$ and $\omega_t(r) = \frac{p_t}{\rho}$ should belong to $(0,1)$ \cite{RRJC} otherwise known as exotic configuration. The radial EoS for various compact star are described graphically in Fig.~\ref{figeos}, which shows linear relationship.\\
We also have calculated the best fit for each EoSs for each stars by using least squares method \cite{GBZGRDD,zdunik}. Fig.~\ref{figfit} describes graphically best fit for each EoSs. The approximation for the best fitted relation for each stars are given as,
\begin{eqnarray}
4U1820-30 &:& p_r ~=~ 0.362175 ~\rho - 157.655 \nonumber \\
SMC~ X-1 &:& p_r ~=~ 0.27393 ~\rho - 101.376 \nonumber \\
SAX~J~1808.4-3658 &:& p_r ~=~ 0.223828 ~\rho - 91.344 \nonumber \\
4U1608-52 &:& p_r ~=~ 0.416461 ~\rho - 183.463 \nonumber \\
PSR~J~1903 +327 &:& p_r ~=~ 0.37333 ~\rho - 153.083 \nonumber \\
Vela~X-1 &:& p_r ~=~ 0.408083 ~\rho - 168.934 \nonumber \\
4U1538-52 &:& p_r ~=~ 0.220122 ~\rho - 90.0515 \nonumber
\end{eqnarray}

\subsection{Moment of inertia and time period}
\begin{figure}[!htbp]
\begin{center}
\begin{tabular}{rl}
\includegraphics[width=7cm]{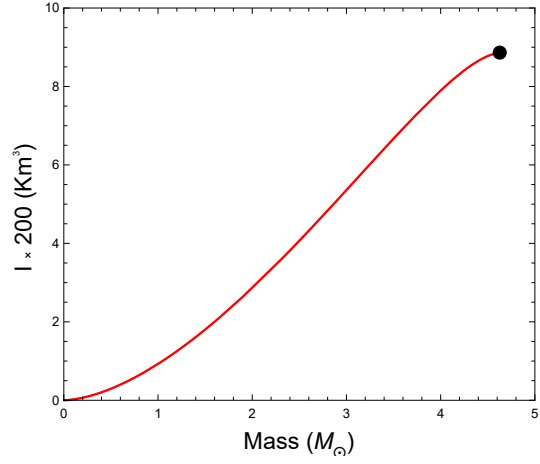}
\\
\end{tabular}
\end{center}
\caption{Moment of inertia with respect to the mass. The solid circle denotes the maximum moment of inertia for the model.}\label{figmi}
\end{figure}

The study of moment of inertia plays a very important role in modeling of compact objects as it allows us to test the stiffness of EoS. The empirical formula for the moment of inertia $I$ transforms a static system to rotating system as suggested by Bejger-Haensel \cite{BH} and it is given by,
\begin{equation}
I = {2 \over 5} (1 + x) M R^2, \label{eqmi}
\end{equation}
where the parameter $x$ is defined as $x = (M/M_\odot)(km/R)$. Here the maximum mass of uniformly slow rotating configuration gives the approximate moment of inertia. The nature of moment of inertia $I$ with respect to mass $M$ is depicted in Fig.~\ref{figmi}.

It can be seen that inertia $I$ is an increasing as the increase of mass and it attains the maximum value for the mass $4.627~ M_\odot$ before declining rapidly. Considering the surface density $\rho(R) = 9.5 \times 10^{14}~gm~cm^{-3}$, we have calculated the $I_{max}$ to be $1773.6$ $km^3$. Comparing the masses of the model star on the Figs.~\ref{figmr} and \ref{figmi}, it can be seen that the mass corresponding to $I_{max}$ is approximately lower by $0.11$\%. \\
This decline of mass indicate the softening of the EoSs without any strong high-density due to hyperonization or phase transition to an exotic state \cite{BBH}.\\
For any non-rotating structure, minimum time-period can be estimated as below provided the EoS obey the sound speeds,
\begin{equation}
\tau \approx 0.82 \sqrt{\left( {M\odot \over M } \right)} \left(\frac{R}{10~ km} \right)^{3 \over 2} ms.
\label{eqtau}
\end{equation}

\begin{figure}[!htbp]
\begin{center}
\begin{tabular}{rl}
\includegraphics[width=7cm]{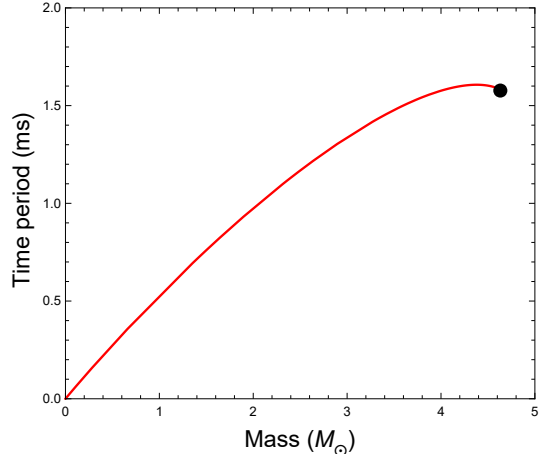}
\\
\end{tabular}
\end{center}
\caption{Variation of time period of rotation with mass.The solid circle denotes the maximum mass of the compact model.}\label{figtau}
\end{figure}
Fig.~\ref{figtau} depicts the variation of time period  with the mass of the model and the maximum time period is obtained to be $1.577~ms$.
\subsection{Mass-central density relationship}
\begin{figure}[!htbp]
\begin{center}
\begin{tabular}{rl}
\includegraphics[width=7cm]{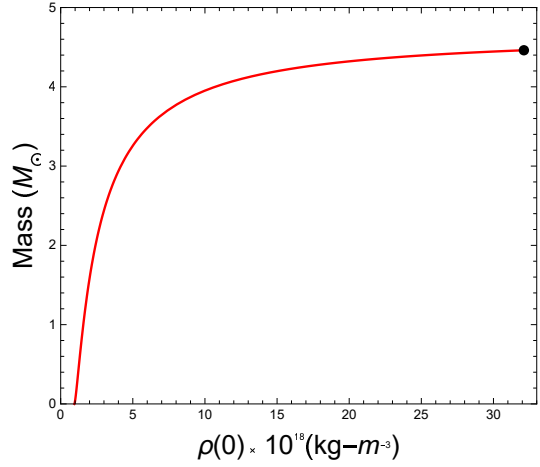}
\\
\end{tabular}
\end{center}
\caption{Variation of central density with the mass. }\label{figmassrho}
\end{figure}

It is evident that the models of cold static compact star represents one parameter family i.e. they can be labeled by using central density or by using central pressure \cite{Haensel}. Fig.~\ref{figmassrho} portrays the profile of mass against the central density and it can be observed that with the increase of the mass of the model, the central density also increases. Moreover, maximum mass is found to be $4.461~M_\odot$ corresponding to the central density $32.11 \times 10^{18}~kg/m^3$.

\subsection{Radius-central density relationship}
\begin{figure}[!htbp]
\begin{center}
\begin{tabular}{rl}
\includegraphics[width=7cm]{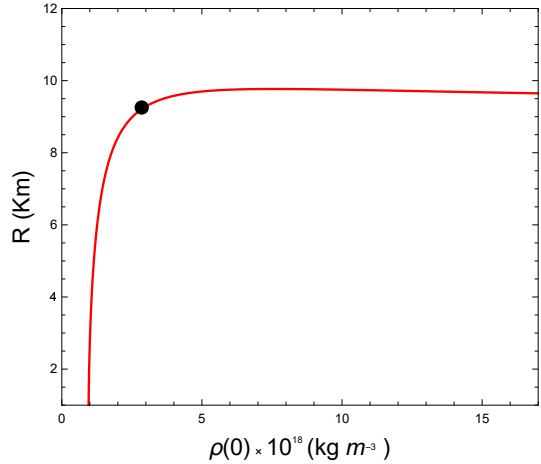}
\\
\end{tabular}
\end{center}
\caption{Variation of central density with the radius. Solid circle denotes the maximum mass of the compact model.}\label{figrcenden}
\end{figure}

The internal structure of any compact model can be studied from its relationship of the radius with any of the matter variable. One such physical test is to study the variation of the central density with the radius of the model as this will allow us to comprehend the nature of the prescribed model. The radius - central density relationship will guide through the process of determining the mass of the compact model and vice-versa. We have studied the nature of the central density with the radius graphically in Fig.~\ref{figrcenden}.
Here the radius increases with the increase of central density until it reaches the critical radius (maximum allowable radius) to remain almost unchanged with the increase of central density. However, the radius for which maximum mass is obtained in Fig.~\ref{figmr}, corresponds to the central density $2.851 \times 10^{18}~kg~m^{-3}$. Additionally the maximum mass occurred in Fig.~\ref{figmassrho} corresponds to the central density $1.11 \times 10^{18}~kg~m^{-3}$.
\subsection{Bounds on model parameter}
Bounds on the parameters $A$, $C$, $D$ are described as,

\begin{center}
\begin{tabular}{ccc}\hline
Conditions & At center $(r=0)$ & At surface $(r = R)$ \\ \hline

$e^\lambda(r)>0$ & satisfied &  $ -\frac{1}{R^2}< A< \frac{2}{R^2} $ \\

$e^\nu(r)>0 $ &  satisfied &  satisfied \\

$\rho(r) >0$ & $A>0$ & $A>0$ \\

$p_r(r)>0$ & $\frac{6 D^2}{C^2} < A < \frac{50 D^2}{3 C^2}$ & $\frac{D}{C}= \frac{\sqrt{A}}{5\sqrt{3(2-A R^2)}}$ \\

$p_t(r)>0$ & $\frac{6 D^2}{C^2} < A < \frac{50 D^2}{3C^2}$ & $0<A<\frac{2}{R^2}$,\\
& & $\frac{3 A}{(2-A R^2)(5+A R^2)}<\frac{D^2}{C^2}$\\
& & $<\frac{A}{3(2-A R^2)}$\\

$\Delta(r)\geq0$ & $0$ & $A>0$ \\

$\frac{d\rho}{dr}\leq0$ & $0$ & $A>0$ \\

Zeldovich's Condition & $\frac{D}{C} \geq \frac{\sqrt{6A}}{7}$ & - \\

$SEC(r)>0$ & $\frac{A}{6}>\frac{D^2}{C^2}$ & $\frac{C^2}{D^2} >\frac{3(2-A R^2)}{A}$ \\

Herrera Condition & $\frac{6 D^2}{C^2}<A<\frac{32 D^2}{3 C^2}$ & same \\

\end{tabular}
\end{center}

We also have calculated $\frac{dp_r}{dr}$, $\frac{dp_t}{dr}$, $\frac{dp_r}{d\rho}$, $\frac{dp_t}{d\rho}$ both at the center and at the surface and combining the above conditions we get the following relations,
\begin{eqnarray}
0~<~A~<~\frac{32}{25 R^2}, \nonumber \\
\frac{25 D^2}{6 C^2}~<~A~<~\frac{32 D^2}{3 C^2}.
\end{eqnarray}
\subsection{Herrera-Ospino-Di Prisco Generating Functions}
Feasible anisotropic solution of EFEs can be obtained by using L. Herrera's \cite{HOP} algorithm. This formalism is essentially an extension of an algorithm proposed by Lake \cite{lake} and it introduces to all solutions with the help of generating functions. More specifically there are two generating functions as suggested by \cite{HOP} to describe all the static spherically symmetric anisotropic fluid matter distribution and the algorithm can be expressed as,
\begin{equation}
e^{\lambda(r)} = \frac{Z^2(r) e^{\int\left( \frac{4}{r^2Z(r)}+2 Z(r) \right) dr}}{r^2\left[ F - 2 \int \frac{Z(r)\left(1 + \Pi(r) r^2 \right) e^{\int\left( \frac{4}{r^2 Z(r)} + 2 Z(r) \right)dr}}{r^8} dr \right]},
\label{gf}
\end{equation}
where $F$ is an arbitrary integrating constant and the corresponding generating functions are as follows,
\begin{eqnarray}
Z(r) &=& {\nu' \over 2} + {1 \over r}, \label{gf1} \\
\Pi(r) &=&  \left(p_r(r) - p_t(r) \right). \label{gf2}
\end{eqnarray}
Here the generating function $Z(r)$ is related to the geometry of the spacetime and $\Pi(r)$ is related to the matter distribution as proposed by Rahaman et. al \cite{RSSP19}. Using Class-I embedding condition in Eq.~(\ref{nu}), the generating functions given in Eqs.~(\ref{gf1})-(\ref{gf2}) take the form,
\begin{eqnarray}
Z(r) &=& {\frac{D~\sqrt{e^{\lambda(r)}-1}}{C + D~\int{\sqrt{e^{\lambda(r)}-1}}dr}} + {1 \over r}, \label{gf3} \\
\Pi(r) &=& \left(p_r(r) - p_t(r) \right). \label{gf4}
\end{eqnarray}

Thus the generating functions to find all exact solutions for our model are given by,
\begin{eqnarray}
Z(r) &=& \frac{\sqrt{3}A D  \sqrt{\frac{A r^2}{2 - A r^2}}}{A C  - \sqrt{3}D \sqrt{\frac{A}{2 - A r^2}}(2 - A r^2)} + {1 \over r}, \label{gf5} \\
\Pi(r) &=& - \Delta(r), \label{gf6}
\end{eqnarray}
where $\Delta(r)$ is given in Eq.~(\ref{aniso}).

\begin{figure}[!htbp]
\centering 
\includegraphics[width=.49\textwidth]{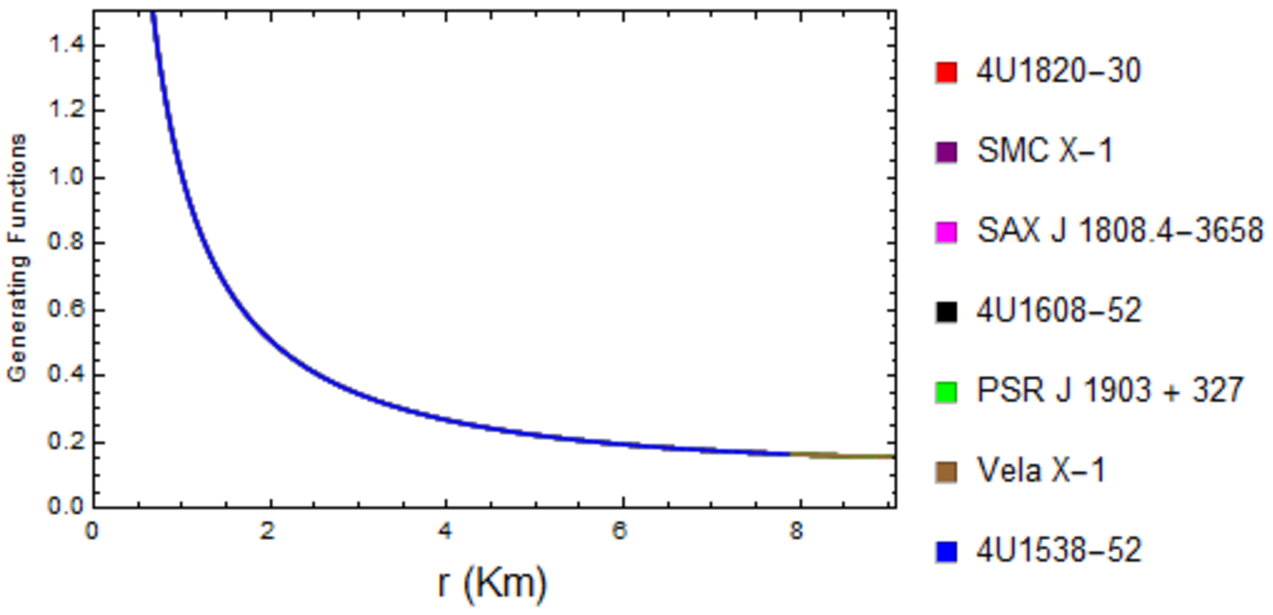}
\hfill
\includegraphics[width=.49\textwidth]{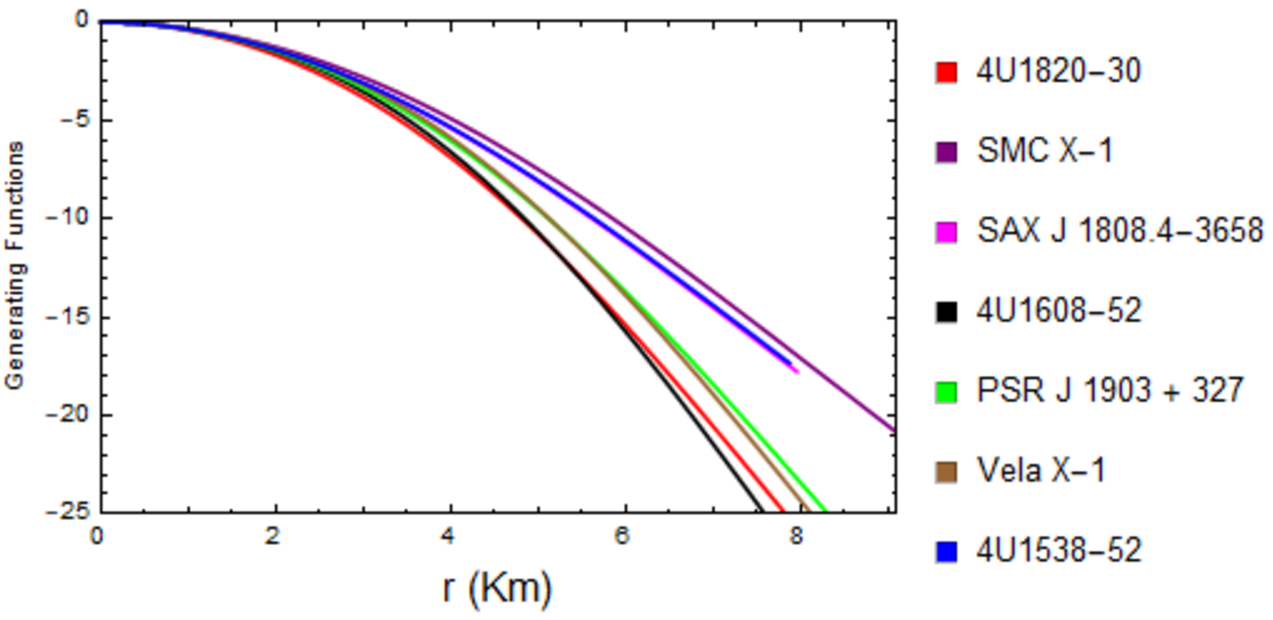}
\caption{\label{figgf} Behavior of the generating functions $Z(r)$ (left side) and $\Pi(r)$ (right side) with respect to the radial coordinate $r$ for several compact stars.}
\end{figure}
Clearly it can be seen in Fig.~\ref{figgf}, the generating function $Z(r)$ related to redshift function is always a positive and decreasing function of `$r$' and the other generating function $\Pi(r)$ is always negative and decreasing in nature.
\section{\label{sec7}Discussions around various compact objects}
To understand our prescribed model we have calculated the parameters for the compact stars $4$U$1820-30$, SMC X$-1$, SAX J $1808.4-3658$, $4$U$1608-52$, PSR J $1903+327$, Vela X$-1$ and $4$U$1538-52$ as given in Table~\ref{table1}. We have predicted the mass and radius of some compact objects using a fixed EoS parameter in Table~\ref{table2}. Further, we have calculated matter variables such as density, radial and transverse sound speed and strong energy conditions for each of the compact stars both at the centers and at the surfaces. The calculated values so obtained are depicted in tabular form in Table~\ref{table3}. Here $|_0$ denotes the value of the matter variable at the center and $|_R$ denotes the same at the boundary of the stars. It can clearly be seen that the value of the matter variables at the surface is smaller than that of the central value, providing a more compact structure. Additionally, the surface redshifts are also provided in Table~\ref{table3} and surface redshifts for each compact stars are within the prescribed range.

\begin{table}[!htbp]
\centering
\setlength{\tabcolsep}{.3\tabcolsep}

\begin{tabular}{|c|c|c|c|c|c|c|c|c|c|c|}\hline
Pulsar  & $\rho|_0$ & $\rho|_R$ & $\frac{dp_r}{d\rho}|_0$ & $\frac{dp_r}{d\rho}|_R$ & $\frac{dp_t}{d\rho}|_0$ & $\frac{dp_t}{d\rho}|_R$ & $SEC|_0$ & $SEC|_R$ & $z|_R$ \\ \hline

$4$U$1820-30$  & $854.09$ & $434.41$ & $0.3589$ & $0.3513$ & $0.3125$ & $0.2478$ & $1308.97$ & $498.19$ & $0.4318$   \\

SMC X-$1$ & $629.73$ & $370.47$ & $0.2683$ & $0.2746$ & $0.2025$  & $0.1795$ & $842.52$ & $412.43$ & $0.3095$  \\

SAX J $1808.4-3658$  & $617.86$  & $409.03$ & $0.2194$ & $0.2275$ & $0.1425$ & $0.1349$ & $758.42$ & $444.55$ & $0.2253$ \\

$4$U$1608-52$  & $918.31$ & $437.95$ & $0.4167$ & $0.3939$ & $0.3821$ & $0.2841$ & $1515.54$ & $508.86$ & $0.4939$ \\

PSR J $1903+327$    & $815.04$  & $408.74$ & $0.3704$ & $0.3601$ & $0.3264$ & $0.2554$ & $1268.52$ & $470.10$ & $0.445$ \\

Vela X-$1$   & $854.63$  & $411.52$ & $0.4072$ & $0.3872$ & $0.3707$ & $0.2785$ & $1394.45$ & $477.24$ & $0.4844$ \\

$4$U$1538-52$     & $612.82$  & $409.97$ & $0.2158$ & $0.2238$ & $0.138$ & $0.1313$ & $747.06$ & $444.63$ & $0.2183$ \\ \hline
\end{tabular}
\caption{\label{table3} Matter variables for different compact objects.}
\end{table}

\section{\label{sec8}Concluding remarks}
In this paper, we have presented a model for spherically symmetric anisotropic sphere considering a specific $e^{\lambda(r)}$. The model is assumed to satisfy Karmarkar condition and smooth matching of interior metric conditions with Schwarzschild exterior solution generate the constants of the model. To analyze our obtained solutions we have considered the physical parameters for some well-known compact stars $4$U$1820-30$, SMC X$-1$, SAX J $1808.4-3658$, $4$U$1608-52$, PSR J $1903+327$, Vela X$-1$ and $4$U$1538-52$ to examine the acceptability of our model.
Some of the key features of our solutions are discussed briefly:\\
\begin{itemize}
\item Both the metric potentials are shown to be regular, well-defined and positive throughout the stellar interior. Moreover, both the metric potentials are described to be finite both at the center and at the boundary to provide an applicable compact model, Fig.~\ref{figmp} supports that.
\item The energy density of a compact star is positive and monotonically decreasing in nature towards the surface with the maximum density at the center making the interior compact. The radial pressure and as well as transverse pressure are monotonically decreasing towards the surface. So all the potentials and parameters are well-valued as illustrated in Figs.~\ref{figpressure} and \ref{figda}. Moreover, the anisotropy is positive as depicted in Fig.~\ref{figda} i.e. the anisotropic force is acting outwards and that leads to model a stable configuration \cite{GM}.
\item For singularity test, we have also computed Kretchsmann Scalar and we found that the model has a singularity at $r =1$ (see Fig.~\ref{figks}), though the model is free from any singularity at the center.
\item All energy conditions are satisfied for this anisotropic physical matter distribution and evidently, we can see from Fig.~\ref{figenergy} that our solutions satisfy all the energy conditions within the star. Additionally it can also be seen that inequality $\rho - p_r - 2 p_t > 0$ is satisfied within the stellar interior.
\item Under the combined action of gravitational, hydrostatic and anisotropic forces our model stays in equilibrium just like any anisotropic fluid configuration needs to be. The behaviour of our model in the effect of TOV equation is shown in Fig.~\ref{figforce}.
\item Stability of our model has been checked using Herrera \cite{herrera92} condition. Also, the absolute value of the difference between radial and transverse sound speeds fulfil Herrera \cite{herrera92} and Andr{\'e}asson \cite{andreasson} criteria making our model potentially stable.
\item Compactness and surface redshifts are examined graphically for our model in Fig.~\ref{figcompred}. It is also examined numerically in Tables~\ref{table2} and \ref{table3}. Additionally, Fig.~\ref{figcompred} depicts that maximum value is attained at the surface and redshifts vanishes at the centre and Table~\ref{table3} shows that maximum value of surface redshifts are $<~5.11$ as suggested by Ivanov \cite{ivanov}.
\item We have plotted mass-radius plot in Fig.~\ref{figmr} and the maximum mass is obtained to be $4.632~M_\odot$ for the radius $9.254$ km. Also Fig.~\ref{figmi} depicts the mass-inertia plot where the maximum mass is approximately $0.11$ \% lower than that in Fig.~\ref{figmr}, which expresses the stiffness of EoS.
\item We have tested the stability of the model by studying mass-central density and radius-central density relationship of the model in the Figs.~\ref{figmassrho}-\ref{figrcenden} respectively. The maximum radius of the model is obtained corresponding to the central density $32.11 \times 10^{18}~kg/m^3$ in Fig.~\ref{figmassrho}. Moreover, Fig.~\ref{figrcenden} suggests that the radius for which the maximum mass is obtained corresponds to the central density $2.851 \times 10^{18}~kg/m^3$.
\end{itemize}
We have tested all the stability criterion both analytically and graphically as depicted in Sec.~\ref{sec6}. Additionally, we have calculated mass and radius for some stars as given in Table.~\ref{table2} taking a certain value of the EoS parameter and one can agree that the predicted radii are in good agreement with the observational data. The generating functions to find all possible exact solutions for our model are also generated. It is shown graphically that $Z(r)$ is always positive and decreasing in nature and $\Pi(r)$ is always negative and increasing in nature which is necessary to be a stable compact star.
Thus it can be concluded that this model satisfies all the properties to be anisotropic compact configurations.

\section*{Acknowledgments}
 FR and SD are thankful to the Inter-University Centre for Astronomy and Astrophysics (IUCAA), Pune, India for providing Visiting Associateship.



\begin{thebibliography}{99}



\bibitem{ks1}
K. Schwarzschild, \emph{{\"U}ber das gravitationsfeld eines massenpunktes nach der Einstein'schen theorie}, \emph{Sitz. Deut. Akad. Wiss}, \emph{Berlin Kl. Math. Phys.} {\bf 1916} (1916) 189. [English translation \emph{On the gravitational field of a point mass, According to Einstein's theory}, \emph{Gen. Relativ. Gravit.} {\bf 35} (2003) 951-959].

\bibitem{ks2}
K. Schwarzschild, \emph{Sitz. Deut. Akad. Wiss}, \emph{Berlin Kl. Math. Phys.} {\bf 24} (1916) 424. [English translation \emph{On the gravitational field of a sphere of incompressible fluid according to Einstein's theory} (1999) arXiv: physics/9912033v1].

\bibitem{tolman}
R. C. Tolman, \emph{Static solutions of Einstein's field equations for spheres of fluid},\emph{Phys., Rev}. {\bf 55} (1939) 364.

\bibitem{OV}
J. R. Oppenheimer and G. M. Volkoff, \emph{On massive neutron cores}, \emph{Phys. Rev.} {\bf 55} (1939) 374.

\bibitem{jeans}
J. Jeans, \emph{The motions of stars in a Kapteyn universe}, \emph{Mon. Not. R. Astron. Soc.} {\bf 82} (1922) 122.

\bibitem{leimatre}
G. Leimatre, \emph{L'Univers en expansion}, \emph{Ann. Soc. Sci. Brux. A} {\bf 53} (1933) 51.

\bibitem{ruderman}
 M. Ruderman, \emph{Pulsars: Structure and Dynamics}, \emph{Annu. Rev. Astron. Atrophys.} {\bf 10} (1972) 427-476.

\bibitem{BL}
R. L. Bowers and E. P. T. Liang, \emph{Anisotropic spheres in General Relativity}, \emph{Astrophys. J.} {\bf 188} (1974) 657.
\bibitem{DG1}
 K. Dev and M. Gleiser, \emph{Anisotropic stars: Exact solutions}, \emph{Gen. Relativ. Gravit.} {\bf 34} (2002) 1793.
\bibitem{DG2}
 K. Dev and M. Gleiser, \emph{Anisotropic stars II: stability}, \emph{Gen. Relativ. Gravit.} {\bf 35} (2003) 1435.
\bibitem{KW}
R. Kippenhahm and A. Weigert, \emph{Stellar Structure and Evolution} Springer, Berlin, (1990).
\bibitem{HS}
 L. Herrera and N. O. Santos, \emph{Local anisotropy in self-gravitating system}, \emph{Phys. Rep.} {\bf 286} (1997) 53.
 \bibitem{sokolov}
A. I. Sokolov, \emph{Phase transitions in a superfluid neutron liquid}, \emph{JETP} {\bf 79} (1980) 1137.
\bibitem{sawyer}
 R. F. Sawyer, \emph{Condensed $\pi$ phase in Neutron-Star matter}, \emph{Phys. Rev. Lett.} {\bf 29} (1972) 382.
\bibitem{weber}
F. Weber, \emph{Pulsars as astrophysical observations for nuclear and particle physics}, Institute of Physics, Bristol, (1999).
\bibitem{migdal}
A. B. Migdal, \emph{Superfluidity and the moments of inertia of nuclei}, \emph{Nucl. Phys.} {\bf 13} (1959) 655.
\bibitem{MK07}
S. D. Maharaj and S. D. K. Komathiraj, \emph{Generalised compact spheres in electric fields}, \emph{Class Quantum Grav.} {\bf 24} (2007) 4513.

\bibitem{RBBU}
 F. Rahaman, P. Bhar, R. Biswas and A. A. Usmani, \emph{Exact interior solutions in $(2 + 1)$ dimensional spacetime}, \emph{ Eur. Phys. J. C.} {\bf 74} (2014) 2845.
\bibitem{bhar1}
 P. Bhar, \emph{Singularity free anisotropic strange quintessence star}, \emph{Astrophys. Space Sci} {\bf 356} (2015) 309.
 \bibitem{bhar2}
P. Bhar, \emph{A new hybrid star model in Krori-Barua spacetime}, \emph{Astrophys. Space Sci} {\bf 357} (2015) 46.
\bibitem{bhar3}
P. Bhar, \emph{Strange star admitting Chaplygin equation of state in Finch-Skea spacetime}, \emph{Astrophys. Space Sci} {\bf 359} (2015) 41.
\bibitem{BM}
P. Bhar and M. H. Murad, \emph{Relativistic compact anisotropic charged stellar models with Chaplygin equation of state}, \emph{Astrophys. Space Sci} {\bf 361} (2006) 334.
\bibitem{TMM}
P. M. Takisa, S. D. Maharaj and C. Mulanga, \emph{Compact relativistic star with quadratic envelope}, \emph{Pramana-J Phys.} {\bf 92} (2019) 40.
\bibitem{herrera92}
 L. Herrera, \emph{Cracking of self-gravitating compact objects}, \emph{Phys. Lett. A},  {\bf 165}, (1992) 206.
\bibitem{HPI}
 L. Herrera, A. Di Prisco and J. Ibanez, \emph{Role of electric charge and cosmological constant in straucture scalars}, \emph{Phys. Rev. D}, {\bf 84}, (2011) 107501.
\bibitem{HOP}
 L. Herrera, J. Ospino and A. Di Parisco, \emph{All static spherically symmetric snisotropic solution of Einstein's equations}, \emph{Phys. Rev. D}, {\bf 77}, (2008) 027502.
 \bibitem{HSW}
L. Herrera, N. O. Santos and A. Wang, \emph{Shearing expansion-free spherical anisotropic fluid evolution}, \emph{Phys. Rev. D},  {\bf 78}, (2008) 084026.
\bibitem{MH03}
M. K. Mak and T. Harko, \emph{Anisotropic stars in general relativity}, \emph{Proc. Roy. Soc. Lond. A}, {\bf 459}, (2003) 393.
\bibitem{HM1}
 T. Harko and M. K. Mak, \emph{Anisotropy in Bianchi-type brane cosmologies}, \emph{Class. Quant. Grav.}, {\bf 21}, (2004) 1489.
\bibitem{HM2}
T. Harko and M. K. Mak, \emph{Anisotropic relativistic stellar models}, \emph{Ann. Phys.(Amsterdam)}, {\bf 11}, (2002) 3.

\bibitem{HM3}
 T. Harko and M. K. Mak, \emph{Quark stars admitting a   one-parameter group of conformal motions}, \emph{Int. J. Mod. Phys. D}, {\bf 13}, (2004) 149.

 \bibitem{HM4}
M. K. Mak and T. Harko, \emph{An exact anisotropic quark star model}, \emph{Chin. J. Astron. Astrophys.},  {\bf 2}, (2002) 248.

\bibitem{MDH}
 M. K. Mak, N. Dobson Jr. and T. Harko, \emph{Exact models for anisotropic relativistic stars}, \emph{Int. J. Mod. Phys. D}, {\bf 11}, (2002) 207.

 \bibitem{MGRD}
S. K. Maurya, Y. K. Gupta, B. Dayanandan and S. Ray, \emph{A new model for spherically symmetric anisotropic compact star}, \emph{Eur. Phys. J. C.}, {\bf 75}, (2016) 225.

\bibitem{MBH}
S. K. Maurya, A. Banerjee and S. Hansraj, \emph{Role of pressure anisotropy on relativistic compact stars}, \emph{Phys. Rev. D}, {\bf 97}, (2018) 044022.

\bibitem{MBG}
S. K. Maurya, A. Banerjee and Y. K. Gupta, \emph{Exact solution of anisotropic compact stars via mass function}, \emph{Astrophys. Space Sci.} {\bf 363}, (2018) 208.

\bibitem{DCRRG}
D. Deb, S. R. Chowdhury, S. Ray, F. Rahaman and B. K.
Guha, \emph{Relativistic model for anisotropic strange stars}, \emph{Annals Phys.}, {\bf 387}, (2017) 239.

\bibitem{KRMH}
M. Kalam, F. Rahaman, S. Molla and S. M. Hossein, \emph{Anisotropic quintessence stars}, \emph{Astrophys. Space Sci.}, {\bf 349}, (2014) 865.

\bibitem{MMKP}
S. K. Maurya, S. D. Maharaj, J. Kumar and A. K. Prasad, \emph{Effect of pressure anisotropy on Buchdahl-type relativistic compact stars}, \emph{Gen. Relativ. and Gravit.}, {\bf 51}, (2019) 86.

\bibitem{TRSD}
 S. Thirukkanesh, F. C. Ragel, R. Sharma and S. Das, \emph{Anisotropic generalization of well-known solutions describing relativistic self-gravitating fluid systems: an algorithm}, \emph{Eur. Phys. J. C.},  {\bf 78}, (2018) 31.

 \bibitem{DRB}
S. Das, F. Rahaman and L. Baskey, \emph{A new class of compact stellar model compatible with observational data}, \emph{Eur. Phys. J. C.} {\bf 79}, (2019) 853.

\bibitem{FS}
T. Feroze and A. Siddiqui, \emph{Charged anisotropic matter with quadratic equation of state}, \emph{Gen. Rel. Grav.}, {\bf 43}, (2011) 1025-1035.

\bibitem{malaver1}
M. Malaver, \emph{Analytical models for compact stars with a linear equation of state}, \emph{World Sci. News}, {\bf 50}, (2016) 64-73.

\bibitem{malaver2}
M. Malaver, \emph{Strange Quark Star Model with Quadratic Equation of State}, \emph{Front. Math. \& Its Appl.}, {\bf 1(1)}, (2014) 9-15.

\bibitem{malaver3}
M. Malaver, \emph{Relativistic Modeling of Quark Stars with Tolman IV Type Potential}, \emph{Int. J Mod. Phys. Appl.}, {\bf 2(1)}, (2015) 1-6.

\bibitem{TM}
P. M. Takisa and S. D. Maharaj, \emph{Some charged polytropic models}, \emph{Gen. Rel. Grav.}, {\bf 45}, (2013) 1951-1969.

\bibitem{TR}
S. Thirukkanesh and F. C. Ragel, \emph{Exact anisotropic sphere with polytropic equation of state}, \emph{PRAMANA Journal of physics}, {\bf 78(5)}, (2012) 687-696.

\bibitem{RS}
L. Randall and R. Sundrum, \emph{An alternative to Compactification}, \emph{Phys. Rev. Lett.}, {\bf 83}, (1999) 3370.

\bibitem{schlai}
 L. Schlaefli, \emph{Sugli spazii di curvatura constante}, \emph{Ann. di Mat.} {\bf 5}, (1871) 170.

\bibitem{janet}
M. Janet, \emph{Sur la possibilit{\'e} de plonger un espace riemannian donn{\'e} dasn un espace euclidien}, \emph{Ann. Soc. Polon. Math.}, {\bf 5}, (1927) 38.

\bibitem{cartan}
{\'E}. Cartan, \emph{Sur la possibilit{\'e} de plonger un espace riemannian donn{\'e} dasn un espace euclidien}, \emph{Ann. Soc. Polon. Math.}, {\bf 6}, (1927) 1.

\bibitem{friedman}
A. Friedman, \emph{Isometric embedding of Riemannian manifolds into Euclidean spaces}, \emph{Rev. Mod. Phys.}, {\bf 37}, (1965) 201.

\bibitem{VT}
 P. C. Vaidya and R. Tikekar, \emph{Exact relativistic model for a superdense star}, \emph{J Astrophys. Aston.}, {\bf 3}, (1982) 325.

\bibitem{tikekar}
R. Tikekar, \emph{Exact model for a relativistic star}, \emph{J Math. Phys.}, {\bf 31}, (1990) 2454.

\bibitem{SPG}
K. N. Singh, N. Pant and M. Govender, \emph{Physical viability of fluid spheres satisfying the Karmarkar condition}, \emph{Eur. Phys. J. C}, {\bf 77}, (2017) 100.

\bibitem{MBJKPP}
S. K. Maurya, A. Banerjee, M. K. Jasim, J. Kumar, A. K. Prasad and A. Pradhan, \emph{Anisotropic compact stars in the Buchdahl model: A comprehensive study}, \emph{Phys. Rev. D}, {\bf 99}, (2019) 044029.

\bibitem{karmarkar}
 K. R. Karmarkar, \emph{Gravitational metrics of spherically symmetry and class one}, \emph{Proc. Ind. Acad. Sci. A}, {\bf 27}, (1948) 56.

 \bibitem{PTGRS}
D. M. Pandya, B. Thakore, R. B. Goti, J. P. Rank and S. Shah, \emph{Anisotropic compact star model satisfying Karmarkar conditions}, \emph{Astrophys. \& Space Sci.}, {\bf 365}, (2020) 1.

\bibitem{PT}
D. M. Pandya and V. O. Thomas, \emph{Models of compact stars of embedding class one for anisotropic distributions satisfying Karmarkar condition}, \emph{Canadian Journal of Phys.}, {\bf 97(3)}, (2019) 337-344.

\bibitem{SBMP}
K. N. Singh, R. K. Bisht, S. K. Maurya and N. Pant, \emph{Static fluid spheres admitting Karmarkar condition}, \emph{Chinese Phys. C},  {\bf 44(3)}, (2020) 035101.

\bibitem{SBLR}
 K. N. Singh, P. Bhar, M. Laishram and F. Rahaman, \emph{A generalised class one static solution}, \emph{Heliyon}, {\bf 5}, (2019) 8.

\bibitem{SERD}
K. N. Singh, A. Errehymy, F. Rahaman and M. Daoud, \emph{Exploring physical properties of compact stars in f(R) gravity: an embedding approach}, (2020) arXiv:2002.08160v1 [gr-qc].

\bibitem{SMERD}
K. N. Singh, S. K. Maurya, A. Errehymy, F. Rahaman and M. Daoud, \emph{Physical properties of class I compact star model for linear and Starobinsky - f(R,T) functions}, \emph{Phys. of Dark Univ.}, {\bf 30}, (2020) 100620.

 \bibitem{GBP1}
S. Gedela, R. K. Bisht and N. Pant, \emph{Relativistic modeling of stellar objects using embedded class one spacetime continuum}, \emph{Mod. Phys. Lett. A}, {\bf 33(1)}, (2020) 2050097.

\bibitem{GBP2}
 S. Gedela, R. K. Bisht and N. Pant, \emph{Relativistic modeling of Vela X$-1$ using the Karmarkar condition}, \emph{Mod. Phys. Lett. A}, {\bf 9}, (2019) 59.

 \bibitem{GPBP}
S. Gedela, R. P. Pant, R. K. Bisht and N. Pant, \emph{A new parametric class of exact solutions of EFEs under the Karmarkar condition for anisotropic fluids}, \emph{Eur. Phys. Journal A}, {\bf 55}, (2019) 95.

\bibitem{OMES}
F. T. Ortiz, S. K. Maurya, A. Errehymy and K. N. Singh, \emph{Anisotropic relativistic fluid spheres: an embedding class I approach}, \emph{Eur. Phys. J. C.},  {\bf 79}, (2019) 885.

\bibitem{JMA}
M. K. Jasim, S. K. Maurya and A. S. M. Al-Sawaii, \emph{A generalised embedding class one static solution describing anisotropic fluid sphere},  \emph{Astrophys. and Space Sc.}, {\bf 365}, (2020) 9.

\bibitem{SSRSS}
N. Sarkar, S. Sarkar, F. Rahaman, K. N. Singh and H. H. Shah, \emph{Anisotropic fluid spheres satisfying the Karmarkar condition}, \emph{Mod. Phys. Lett. A} {\bf 34(15)}, (2019) 1950113.

\bibitem{SSSR} N. Sarkar, K. N. Singh, S. Sarkar and F. Rahaman, \emph{Compact star model in class I spacetime}, \emph{Eur. Phys. J. C.}, {\bf 79}, (2019) 516.

\bibitem{PKMD}
A. K. Prasad, J. Kumar, S. K. Maurya and B. Dayanandan, \emph{Relativistic model for anisotropic compact stars using Karmarkar condition}, \emph{Astrophys. Space Sci.}, {\bf 364}, (2019) 66.

\bibitem{FMS}
S. Fatema, M. H. Murad and K. N. Singh, \emph{New exact anisotropic static spherically symmetric stellar models satisfying the Eiesland condition}, \emph{Ann. Phys.}, {\bf 402}, (2019) 1.

\bibitem{TF}
R. Tamta, P. Fuloria, \emph{Analysis of physically realizable stellar models in embedded class one spacetime manifold}, \emph{Mod. Phys. Lett. A}, {\bf 35(04)}, (2020) 2050001.

\bibitem{SF}
M. F. Shamir, I. Fayyaz, \emph{Effect of f(R) gravity modes on compact stars}, \emph{Th. Math. Phys.}, {\bf 202}, (2020) 112.

\bibitem{RSERD}
M. Rahaman, K. N. Singh, A. Errehymy, F. Rahaman and M. Daoud, \emph{Anisotropic Karmarkar stars in f(R,T) gravity}, \emph{Eur. Phys. J. C.}, {\bf 80}, (2020) 272.

\bibitem{MOJ}
S. K. Maurya, Francisco Tello-Ortiz and M. K. Jasim, \emph{An EGD model in the background of embedding class I space–time}, \emph{Eur. Phys. J. C}, {\bf 80}, (2020) 918.

\bibitem{OML}
Francisco Tello-Ortiz, S. K. Maurya and Y. Gomez-Leyton, \emph{Class I approach as MGD generator}, Eur. Phys. J. C, {\bf 80}, (2020) 324.

\bibitem{GMSP}
M. Govender, A. Maharaj, K. N. Singh and N. Pant, \emph{Dissipative collapse of a Karmarkar star}, \emph{Mod. Phys. Lett. A} (2020) https://doi.org/10.1142/S0217732320501643.

\bibitem{KCN}
M. Kohler and K. L. Chao, \emph{Zentralsymmetrische statische Schwerefelder mit R{\"a}umen der Klasse $1$}, \emph{Z. Naturforsch. A}, {\bf 20}, (1965) 1537.

\bibitem{MGTR}
S. K. Maurya, Y. K. Gupta, Smitha T. T. and F. Rahaman, \emph{A new exact anisotropic solution of embedding class one},   \emph{ Eur.Phys.J.A. } { \bf52 } (2016)  191.

\bibitem{buchdahl59}
 H. A. Buchdahl, \emph{General relativistic fluid spheres}, \emph{Phys. Rev.}, {\bf 116}, (1959) 1027.

\bibitem{DL}
M. S. R. Delgaty and K. Lake, \emph{Physical acceptability of isolated, static, spherically symmetric, perfect fluid solutions of Einstein's equations}, \emph{Comput. Phys. Commun.}, {\bf 115}, (1998) 395.

\bibitem{gauss}
C. F. Gauss, \emph{ Disquisitiones Generales circa Supercies Curvas} (1827).

\bibitem{codazzi}
D. Codazzi, \emph{Sulle coordinate curvilinee d'una superficie e dello spazio}, \emph{Ann. di Mat}, {\bf 2}, (1868) 269.

\bibitem{kasner}
E. Kasner, \emph{Finite representation of the solar gravitational field in flat space of six dimensions}, \emph{Am. J. Math.}, {\bf 43}, (1921) 130.

\bibitem{GG}
Y. K. Gupta and M. P. Goyel, \emph{Class two analogues of TY Thomas's theorem and different types of embeddings of static spherically symmetric space-times}, \emph{Gen. Relativ. Gravit.}, {\bf 6}, (1975) 499.

\bibitem{eddington}
A. S. Eddington Kasner, \emph{The Mathematical Theory of Relativity} Cambridge University Press, Cambridge, (1924).

\bibitem{PS}
 S. N. Pandey and S. P. Sharma, \emph{Insufficiency of Karmarkar's condition}, \emph{Gen. Relativ. Gravit.},  {\bf 14}, (1981) 113.

\bibitem{MGRC} S. K. Maurya, Y. K. Gupta, S. Ray and S. R. Chowdhury, \emph{Spherically symmetric charged compact stars}, \emph{Eur. Phys. J. C}, {\bf 75}, (2015) 389.

\bibitem{MGDJA}
 S. K. Maurya, Y. K. Gupta , B. Dayanandan, M. K. Jasim and A. Al-Jamel, \emph{Relativistic anisotropic models for compact star with equation of state $p = f(\rho)$}, \emph{Int. Journal of Mod. Phys. D}, {\bf 26}, (2017) 1750002.

\bibitem{lake}
 K. Lake, \emph{All static spherically symmetric perfect-fluid solutions of Einstein's equations}, \emph{Phys. Rev. D}, {\bf 67}, (2003) 104015.

\bibitem{MK}
R. Mansouri and M. Khaorrami, \emph{The equivalence of Darmois-Israel and distributional method for thin shells in general relativity}, \emph{J. Math. Phys.}, {\bf 37}, (1996) 5672.

\bibitem{israel}
W. Israel, \emph{Singular hypersurfaces and thin shells in general relativity}, \emph{Nuovo Cimento}, {\bf 44B}, (1966) 1; corrections in {\bf 48 B}, (1966) 463.

\bibitem{darmois}
G. Darmois, \emph{Les equations de la gravitation einsteinienne}, Chapitre V, Memorial de Sciences Mathematiques, Fascicule XXV (1927).

\bibitem{MS}
 C. W. Misner and D. H. Sharpe, \emph{Relativistic equations for adiabatic, spherically symmetric gravitational collapse}, \emph{Phys. Rev.}, {\bf 136}, (1964) B571.

\bibitem{GRDDD}
 T. Gangopadhyay, S. Ray, X-D. Li, J. Dey and M. Dey, \emph{Strange star equation of state fits the refined mass measurement of $12$ pulsars and predicts their radii}, \emph{Mon. Not. R. Astron. Soc.}, {\bf 431}, (2013) 3216.

\bibitem{RCKSR}
 F. Rahaman, K. Chakraborty, P. K. F. Kuhttig, G. C. Shit and M. Rahaman, \emph{A new deterministic model of strange stars}, \emph{Eur. Phys. J. C.}, {\bf 74}, (2014) 3126.

\bibitem{elebert}
 P. Elebert et al, \emph{Optical spectroscopy and photometry of SAX J$1808.4-3658$ in outburst}, \emph{Mon. Not. Roy. Astron. Soc.}, {\bf 395}, (2009)  884.
\bibitem{GOCW}
T. Guver, F. Ozel, A. Cabrera-Lavers and P. Wroblewski, \emph{The distance, mass and radius of the neutron star in $4$U $1608-52$}, \emph{Astrophys. J.}, {\bf 712}, (2010) 964.

\bibitem{freire}
P. C. C. Freire et al, \emph{On the nature and evolution of the unique binary pulsar J$1903+0327$}, \emph{Mon. Not. Roy. Astron. Soc.}, {\bf 412}, (2011) 2763.

\bibitem{rawl}
M. L. Rawl, J. A. Orosz, J. E. McClintock, M. AP. Torres, C. D. Bayin, M. M. Buxton, \emph{Refined neutron star mass determinations for six eclipsing x-ray pulsar binaries}, \emph{Astrophys. J.}, {\bf 730}, (2011) 25.

\bibitem{leibovitz}
 C. Leibovitz, \emph{Spherically symmetric static solutions of Einstein's equations}, \emph{Phys. Rev. D}, {\bf 185}, (1969) 1664.

\bibitem{PMP}
 N. Pant, R. N. Mehta and M. J. Pant, \emph{New class of regular and well behaved exact solutions in general relativity}, \emph{Astrophys. Space Sci.}, {\bf 330}, (2010) 335.

\bibitem{pant}
N. Pant, \emph{Some new exact solutions with finite central parameters and uniform radial motion of sound}, \emph{Astrophys. Space Sci.}, {\bf 331}, (2011) 633.

\bibitem{GM}
M. K. Gokhroo and A. L. Mehra, \emph{Anisotropic spheres with variable energy density in general relativity}, \emph{Gen. Relativ. Gravit.}, {\bf 26} (1994) 75.

\bibitem{ZN}
 Y. B. Zeldovich and I. D. Novikov, \emph{Relativistic Astrophysics, Vol. 1, Stars and Relativity}, The University of Chicago Press, Chicago, (1971).

\bibitem{varela}
V. Varela, F. Rahaman, S. Ray, K. Chakraborty and M. Kalam, \emph{Charged anisotropic matter with linear or nonlinear equation of state}, \emph{Phys. Rev. D} {\bf 82}, (2010) 044052.

\bibitem{ponce}
J. Ponce de Le{\'o}n, \emph{Limiting configurations allowed by the energy conditions}, \emph{Gen. Relativ. Gravity} {\bf 25}, (1993) 1123.

\bibitem{MESTD}
S. K. Maurya, A. Errehymy, K. N. Singh, F. Tello-Ortiz and M. Daoud, \emph{Gravitational decoupling minimal geometric deformation model in modified $f(R,T)$ gravity theory}, \emph{Phys of Dark Univ.}, {\bf 30}, 100640 (2020).

\bibitem{HE}
S. W. Hawking, G. R. Ellis, \emph{The large scale structure of space-time}, (Cambridge Monographs on Mathematical Physics), Cambridge University Press, Cambridge (1973).

\bibitem{NKG}
N. K. Glendenning, \emph{Compact Stars: Nuclear physics, particle physics and general relativity}, Springer, USA (1997).

\bibitem{abreu}
 H. Abreu, H. Hern{\'a}ndez and L. A. N{\'u}{\~n}ez, \emph{Sound speeds, cracking and the stability of self-gravitating anisotropic compact objects}, \emph{Class. Quantum Gravity}, {\bf 24}, (2007) 4631.

\bibitem{HH}
H. Heintzmann and W. Hillebrandt, \emph{Neutron stars with an anisotropic equation of state: Mass, redshift and stability}, \emph{Astron. Astrophys.}, {\bf 38}, (1975) 51.

\bibitem{bondi}
 H. Bondi, \emph{The contraction of gravitating spheres}, \emph{Proc. R. Soc. London A}, {\bf 281},(1987) 39.

\bibitem{chan}
R. Chan, L. Herrera and N. O. Santos, \emph{Dynamical instability for radiating anisotropic collapse},  \emph{Mon. Not. R. Astron. Soc.}, {\bf 265},(1993) 533.

\bibitem{knutsen}
H. Knutsen, \emph{On the stability and physical properties of an exact relativistic model for a superdense star}, \emph{Mon. Not. R. Astron. Soc.}, {\bf 232}, (1988) 163.

\bibitem{HPY}
P. Haensel, A. Y. Potekhin, D. G. Yakovlev, \emph{Neutron Stars I: equation of state and structures}, Springer (2007).

\bibitem{chandrasekhar}
S. Chandrasekhar, \emph{A general variational principle governing the radial and non-radial oscillations of gaseous masses}, \emph{Astrophys. J.}, {\bf 139}, (1964) 664.

\bibitem{harrison}
B. K. Harrison, K. S. Throne, M. Wakano, J. A. Wheeler. \emph{Gravitational Theory and Gravitational Collapse} University of Chicago Press, Chicago, (1965).

\bibitem{ST}
S. Shapiro and S. Teukolsky, \emph{Black Holes, White Dwarfs and Neutron Stars} Wiley, New York (1983).

\bibitem{HBPP}
S. H. Hendi, G. H. Bordbar, B. Eslam Panah and S. Panahiyan, \emph{Neutron stars structure in the context of massive gravity}, arXiv:1701.01039[gr-qc] (2017).

\bibitem{SM}
 R. Sharma and S. D. Maharaj, \emph{A class of relativistic stars with the linear equation of state},\emph{ Mon. Not. R.Astron. Soc.}, {\bf 375}, (2007) 1265.

\bibitem{ThM}
 S. Thirukkanesh and S. D. Maharaj, \emph{ Charged anisotropic matter with linear equation of state}, \emph{ Class. Quant. Grav.}, {\bf 25}, (2008) 23.

\bibitem{RPSD}
B. S. Ratanpal, D. M. Pandya, R. Sharma and S. Das, \emph{Charged compact stellar model in Finch-Skea spacetime}, \emph{Astrophys. Space Sci.}, {\bf 362}, (2017) 82.

\bibitem{RR}
C. E. Rhoades and R. Ruffini, \emph{Maximum Mass of a Neutron Star}, \emph{Phys. Rev. Lett.}, {\bf 32}, (1974) 324.

\bibitem{JT}
K. Jotania and R. Tikekar, \emph{Paraboloidal space-time and relativistic models of strange stars}, \emph{Int. J. Mod. Phys. D}, {\bf 15(08)}, (2006) 1175.

\bibitem{BHamity}
D. E. Barraco and V. H. Hamity, \emph{Maximum mass of a spheically symmetric isotropic star}, \emph{Phys. Rev. D.}, {\bf 65}, (2002) 124028.

\bibitem{bohmer}
C. G. B{\"o}hmer and T. Harko, \emph{Bounds on the basic physical parameters for anisotropic compact general relativistic objects}, \emph{Class. Quan. Grav.}, {\bf 23}, (2006) 6479.

\bibitem{ivanov}
B. V. Ivanov, \emph{Maximum bounds on the surface redshift of anisotropic stars}, \emph{Phys. Rev. D}, {\bf 65}, (2002) 104011.

\bibitem{HBPSC}
A. Hewish, S. J. Bell, J. D. H. Pilkington, P. F. Scott and R. A. Collins, \emph{Observation of a rapidly pulsating radio source}, \emph{Nature}, {\bf 217}, (1968) 709.

\bibitem{FO}
 J. A. Frieman and A. Olinto, \emph{Is the sub-millisecond pulsar strange ?}, \emph{Nature}, {\bf 341}, (1989) 633.

\bibitem{HZ}
 P. Haensel and J. L. Zdunik, \emph{A submillisecond pulsar and the equation of state of dense matter}, \emph{Nature}, {\bf 340}, (1989) 617.

\bibitem{PBP}
 M. Prakash, E. Baron and M. Prakash, \emph{Rotation of stars containing strange quark matter}, \emph{Phys. Lett. B}, {\bf 243}, (1990) 175.

\bibitem{LPMY}
J. M. Lattimer, M. Prakash, D. Masak and A. Yahil, \emph{Rapidly rotating pulsars and the equation of state}, \emph{Astrophys. J}, {\bf 355}, (1990) 241.

\bibitem{zdunik}
J. L. Zdunik, \emph{Strange stars- linear approximation of the EoS and maximum QPO frequency}, \emph{Astron. Astrophys.}, {\bf 359}, (2000) 311.

\bibitem{HC}
T. Harko and K. S. Cheng, \emph{Maximum mass and radius of strange stars in the linear approximation of the EoS}, \emph{Astron. Astrophys.}, {\bf 385}, (2002) 947.

\bibitem{DBDRS}
M. Dey, I. Bombacci, J. Dey, S. Ray and B. C. Samanta, \emph{Strange Stars with realistic quark
vector interaction and phenomenological density-dependent scalar potential}, \emph{Phys. Lett. B},{\bf 438}, (1998) 123.

\bibitem{GBZGRDD}
D. Gondek-Rosinska, T. Bulik, L. Zdunik, E. Gourgoulhon, S. Ray, J. Dey and M. Dey,
\emph{Rotating compact strange stars}, \emph{Astron. Astrophys.}, {\bf 363}, (2000) 1005.

\bibitem{RRJC}
F. Rahaman, S. Ray, A. K. Jafry and K. Chakraborty, \emph{Singularity-free solutions for anisotropic charged fluids with Chaplygin equation of state}, \emph{Phys. Rev. D}, {\bf 82}, (2010) 104055.

\bibitem{BH}
M. Bejger and P. Haensel, \emph{Moments of inertia for neutron and strange stars: limits derived for the Crab pulsar}, \emph{Astron. Astrophys.}, {\bf 396}, (2002) 3.

\bibitem{BBH}
M. Bejger, T. Bulik and P. Haensel, \emph{Constraints on the dense matter equation of state from the measurements of PSR J $0737 - 3039$ A moment of inertia and PSR J $0751 + 1807$ mass}, \emph{Mon. Not. R. Astron. Soc.}, {\bf 364}, (2005) 635.

\bibitem{Haensel}
P. Haensel, \emph{Equation of state of dense matter and maximum mass of neutron stars}, \emph{Final Stages of Stellar Evolution edts. J.-M. Hameury \& C. Motch EAS Publications Series}, (2008).

\bibitem{RSSP19}
F. Rahaman, S. Sarkar, K. N. Singh and N. Pant, \emph{Generating functions of wormholes}, \emph{Mod. Phys. Lett. A}, {\bf 34}, (2019) 1950010.

\bibitem{andreasson}
H. Andr{\'e}asson, \emph{Sharp bounds on the critical stability radius for relativistic charged spheres}, \emph{Commun. Math. Phys.} {\bf 288},  (2009) 715.

\end{thebibliography}
\end{document}